\documentclass[reprint,secnumarabic,amssymb, nobibnotes, aps, prl, superscriptaddress, abstract, greek, floatfix]{revtex4-2}

\usepackage{graphicx}
\usepackage{xfrac}
\usepackage{appendix}
\usepackage{upgreek}

\begin{document}

\title{Cryogenic growth of tantalum thin films for low-loss superconducting circuits}

\author{Teun A. J. van Schijndel}
\email[]{teunvanschijndel@ucsb.edu}
\affiliation{Electrical and Computer Engineering Department, University of California Santa Barbara, Santa Barbara, California 93106, USA}

\author{Anthony P. McFadden}
\affiliation{National Institute Of Standards and Technology, Boulder, Colorado 80305, USA\looseness=-1}

\author{Aaron N. Engel}
\affiliation{Materials Department, University of California Santa Barbara, Santa Barbara, California 93106,  USA\looseness=-1}

\author{Jason T. Dong}
\affiliation{Materials Department, University of California Santa Barbara, Santa Barbara, California 93106, USA\looseness=-1} 

\author{Wilson J. Y\'{a}nez-Parre\~{n}o}
\affiliation{Electrical and Computer Engineering Department, University of California Santa Barbara, Santa Barbara, California 93106, USA}

\author{Manisha Parthasarathy}
\affiliation{Rutgers University, New Brunswick, New Jersey 08901, USA}

\author{Raymond W. Simmonds}
\affiliation{National Institute Of Standards and Technology, Boulder, Colorado 80305, USA\looseness=-1}

\author{Christopher J. Palmstr\o{}m}%
\email[]{cjpalm@ucsb.edu}
\affiliation{Electrical and Computer Engineering Department, University of California Santa Barbara, Santa Barbara, California 93106, USA}
\affiliation{Materials Department, University of California Santa Barbara, Santa Barbara, California 93106, USA\looseness=-1}

\date{\today}%

\begin{abstract}
Motivated by recent advancements highlighting Ta as a promising material in low-loss superconducting circuits and showing long coherence times in superconducting qubits, we have explored the effect of cryogenic temperatures on the growth of Ta and its integration in superconducting circuits. Cryogenic growth of Ta using a low temperature molecular beam epitaxy (MBE) system is found to stabilize single phase $\upalpha$-Ta on several different substrates, which include Al\textsubscript{2}O\textsubscript{3}(0001), Si(001), Si(111), SiN\textsubscript{x}, and GaAs(001). The substrates are actively cooled down to cryogenic temperatures and remain $<$ 20 K during the Ta deposition. X-ray $\uptheta$-$2\uptheta$ diffraction after warming to room temperature indicates the formation of polycrystalline $\upalpha$-Ta. The 50 nm $\upalpha$-Ta films grown on Al\textsubscript{2}O\textsubscript{3}(0001) at a substrate manipulator temperature of 7 K have a room temperature resistivity ($\uprho$\textsubscript{300 K}) of 13.4 $\upmu \Omega$\textit{·}cm, a residual resistivity ratio (RRR) of 17.3 and a superconducting transition temperature (T\textsubscript{C}) of 4.14 K, which are comparable to bulk values. In addition, atomic force microscopy (AFM) indicates that the film grown at 7 K with an RMS roughness of 0.45 nm was significantly smoother than the one grown at room temperature. Similar properties are found for films grown on other substrates. Results for films grown at higher substrate manipulator temperatures show higher $\uprho$\textsubscript{300 K}, lower RRR and Tc, and increased $\upbeta$-Ta content. Coplanar waveguide resonators with a gap width of 3 $\upmu$m fabricated from cryogenically grown Ta on Si(111) and Al\textsubscript{2}O\textsubscript{3}(0001) show low power Q\textsubscript{i} of 1.9 million and 0.7 million, respectively, indicating polycrystalline $\upalpha$-Ta films may be promising for superconducting qubit applications even though they are not fully epitaxial. 

\end{abstract}
\maketitle

\section{Introduction}
Recent advances in superconducting quantum information systems highlight the effectiveness of tantalum as a superconducting material, with quantum bits reaching coherence times up to 0.5 ms \cite{place2021new, wang2022towards}. However, challenges persist as various sources of energy losses arise from material-related factors, consequently limiting further enhancements in qubit performance  \cite{wang2015surface, dial2016bulk, gambetta2016investigating, martinis2005decoherence, pappas2011two, lisenfeld2019electric, wang2014measurement}. Addressing these challenges relies on the optimization of bulk substrate, surface passivation, and the deposition process, to potentially mitigate losses. Crucially, Ta needs to be grown in a body-centered-cubic (BCC) lattice structure ($\upalpha$-Ta), the stable phase in bulk, for the realization of desirable superconducting properties. For example, bulk $\upalpha$-Ta has a superconducting transition temperature (T\textsubscript{C}) of 4.5 K whereas the superconducting phase transition temperature of $\upbeta$-Ta is reported between 0.6 K and 1 K \cite{mazin2022superconducting}. Further, the high chemical stability of Ta allows for aggressive chemical cleaning to remove surface contamination during device fabrication \cite{place2021new, crowley2023disentangling}. It has also been suggested that the inherent properties of its native oxide are superior to those of other superconductors such as Nb, Al, and, TiN \cite{mclellan2023chemical}.

To achieve high-performance qubit devices on sapphire, the substrate is typically annealed \textit{in-situ} in UHV conditions to temperatures ranging from 800-1100 °C to achieve a clean surface before growth of $\upalpha$-Ta thin films \cite{place2021new, wang2022towards, crowley2023disentangling}. Epitaxial $\upalpha$-Ta growth is typically achieved with a substrate temperature of 500 °C \cite{place2021new}. The high substrate temperature during Ta deposition is required as the metastable $\upbeta$-phase is usually reported in thin films obtained from deposition at room temperature \cite{read1965new, westwood1973effects, feinstein1973factors}. Colin \textit{et al.} \cite{colin2017origin}  argue from comparing bulk, interface, and surface energies that at room temperature the metastable $\upbeta$-Ta phase preferentially nucleates over the equilibrium $\upalpha$-Ta phase due to its predicted lower nucleation energy barrier. At growth temperatures above 300 °C the $\upalpha$-Ta is preferentially formed \cite{matson2000effect, gladczuk2004tantalum, zhou2009effects}. Due to its high stability, sapphire allows the growth of epitaxial $\upalpha$-Ta at elevated temperatures without decomposition and detrimental interfacial reactions \cite{gladczuk2004tantalum, zhou2023epitaxial}.

The majority of tantalum qubits are grown on sapphire substrates, but there are reports of low-loss superconducting circuits on Silicon \cite{shi2022tantalum, lozano2022manufacturing}. There would be substantial benefits if silicon substrates could be used as it may be possible to take advantage of the mature silicon technology, which could help to enable scaling of complex systems of many qubits. $\upalpha$-Ta can be grown on Si by heating the substrate similar to sapphire. However, elevated temperatures are likely to lead to Si-Ta interfacial reactions and Ta-silicide formation, which could have detrimental effects on qubit coherence times. Another approach is to reduce the growth temperature by using a nucleation layer, such as Nb \cite{face1987nucleation,sajovec1992structural,urade2024microwave}, Ti \cite{chen2001growth}, Cr \cite{zhang2003formation},
TiN \cite{wu2023high}, or TaN \cite{bernoulli2013magnetron,gladczuk2005sputter,wang2016influence,tsao2013tantalum}. However, this introduces an additional layer that could react with the substrate or metal film, which potentially could have a higher loss tangent due to suppression of superconductivity or additional disorder near the superconductor-substrate interface \cite{alegria2023two, muller2019towards}.

In this study, we demonstrate cryogenic growth using a low-temperature molecular beam epitaxy (MBE) system to stabilize $\upalpha$-Ta independent of the substrate, without using a seed layer, enabling its integration with various material systems. Substrates are actively cooled to cryogenic temperatures and remain below 20 K during the Ta deposition. At these temperatures, it is expected that atom mobility and interfacial reactions should be minimized. This approach holds the potential to enhance the superconducting properties of $\upalpha$-Ta thin films deposited on both sapphire and silicon substrates and also provides an avenue for exploring the use of other substrates. We explore the effect of growth temperature and find that the crystal structure of the Ta formed after low temperature growth and warming to room temperature consists of $\upalpha$-Ta and is polycrystalline. The stabilization of $\upalpha$-Ta is independent of the substrate and the same superconducting properties have been found on both amorphous and crystalline substrates with different orientations. Ultimately, this work describes a novel technique for the development of high-quality superconducting materials and its integration with specific material systems that can potentially improve device performance not only in superconducting circuits but also in elevating capabilities across diverse fields reliant on superconducting heterostructures.

\section{Methods}
Various substrates, including Al$_2$O$_3$(0001), Si(111) and (001), GaAs(001), and amorphous SiN$_x$ deposited on Si(001) were used in this study. The details of the surface preparation are given in Appendix A. The Ta films were deposited in a Scienta Omicron LT EVO 50 MBE chamber with an LN$_2$ shroud holding a base pressure of $2 \times 10^{-11}$ mbar. Ta was evaporated from an electron beam evaporation source in an additional chamber separated by a gate valve. Ultra-high vacuum conditions remain during Ta deposition, with the main chamber pressure staying below $1 \times 10^{-10}$ mbar. The cryogenic manipulator allows for temperatures ranging from 300 K down to 6 K during growth. The temperature is measured near the substrate with a Si diode. The actual substrate temperature is believed to be less than 10 K warmer as determined by attaching a thermocouple on a dummy block with the radiation from a cell at 1000 °C impinging on the dummy block. Growth rate and thickness were monitored using a Quartz-Crystal-Microbalance, targeting a nominal thickness of 50 nm for all films investigated in this study.

DC electrical properties were measured with a 4-point resistance geometry using a Quantum Design Physical Properties Measurement System. Ti/Au contact pads were deposited via \textit{ex-situ} electron beam evaporation, using a shadow mask to avoid lithography and lift-off processes. The crystallography and surface morphology of the thin films were investigated through \textit{ex-situ} X-ray diffraction (XRD) measurements and Atomic Force Microscopy (AFM), respectively. 

To investigate the microwave performance relevant to superconducting electronics applications, films were patterned into coplanar waveguide (CPW) resonators. The design employed consists of eight $\sfrac{1}{4}$-wave, inductively coupled hanger resonators with one central feedline on each chip. This design is similar to that presented in Kopas \textit{et al.} \cite{kopas2022simple}, but modified slightly to obtain the desired resonant frequencies between 5.5-6.5 GHz and external quality factors of Q\textsubscript{c} $\sim $ $2 \times 10^5$. The dimensions of the center conductor (gap) are chosen to be 6 $\upmu$m (3  $\upmu$m). These small dimensions were intentionally chosen to increase the electric field participation of the surfaces and interfaces which typically results in a lower measured Q\textsubscript{i}, but is a more rigorous test of material surfaces and interfaces that typically dominate RF loss at low power \cite{kopas2022simple}.

Following material growth, 10 mm $\times$ 10 mm chips were patterned using direct write photolithography, and the Ta film was etched in an inductively coupled plasma (ICP) tool using Cl\textsubscript{2}/BCl\textsubscript{3}. The same process was used for samples on silicon and sapphire substrates. This etch process will not appreciably etch sapphire, but will etch silicon. Consequently, samples grown on Si substrates are ‘trenched’ with an approximate depth of 100 nm. We note that trenching reduces the electric field participation of the Si surface, Ta/Si interface, and Si substrate in the resonator structures which tends to result in a higher measured Q\textsubscript{i} \cite{calusine2018analysis}.

After processing, cleaned and patterned chips were coated with protective photoresist and diced down to 7.5 mm x 7.5 mm for packaging. After the protective resist was stripped, chips were dipped in 6:1 buffered oxide etch (BOE) for 2 minutes followed by a rinse in DI water and N\textsubscript{2} blow dry. Chips were then wire-bonded with aluminum wire and packaged in a 2-port vetted package for low temperature RF measurements. S\textsubscript{21} (transmission) measurements were performed in a dilution refrigerator having a base temperature of 35 mK. The attenuation and amplification setup used is similar to that presented in \cite{mcrae2020materials}, with the addition of a Josephson parametric amplifier (JPA) installed on the mixing chamber that was used for low power measurements. Resonators were measured at base temperature with varying drive powers typically corresponding to a few million to $\sim$ 0.01 photons in the cavity. 
\begin{figure*}[t!]
    \centering
        \centering        \includegraphics{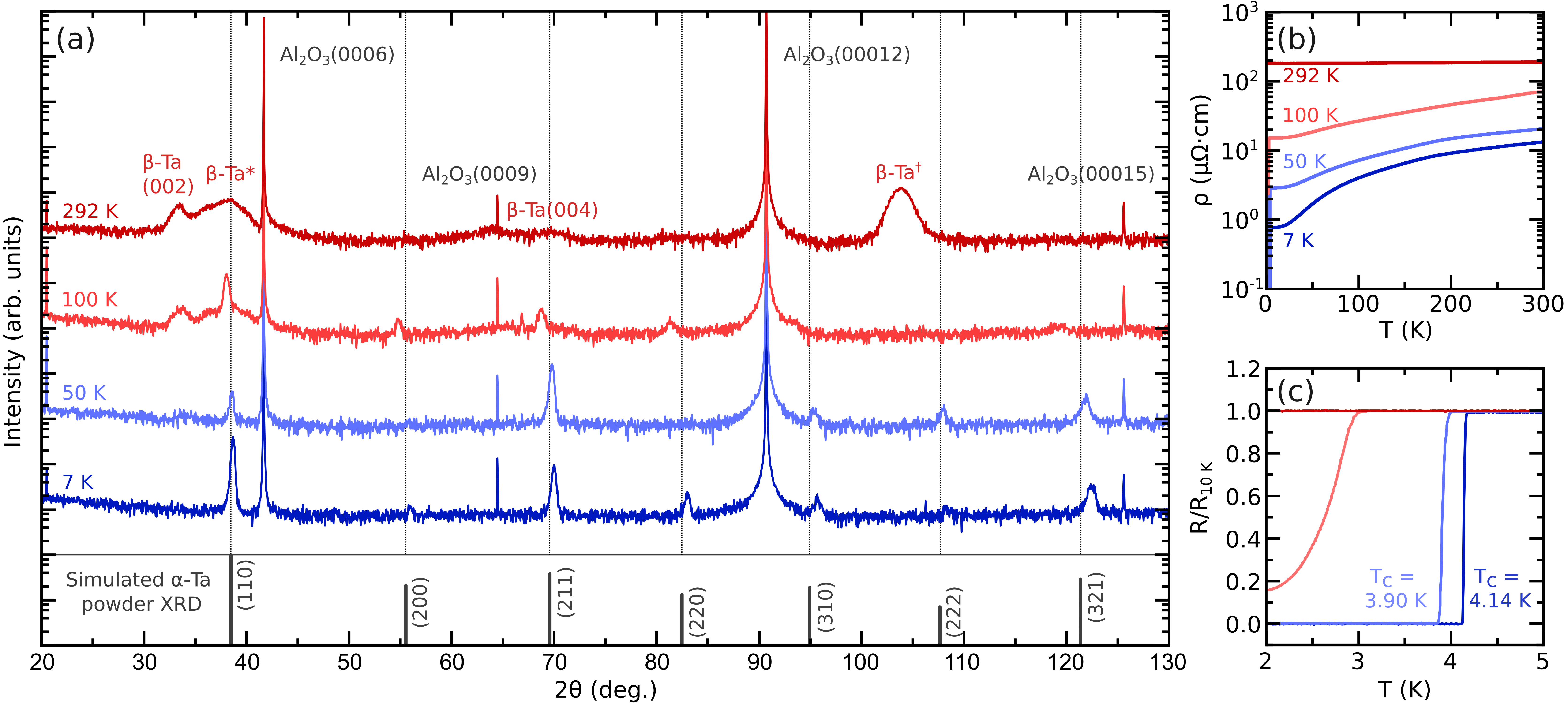} 
        \caption{Structural and electrical characterization of tantalum as a function of growth temperature.  (a) X-ray diffraction pattern for Ta films grown at different temperatures and compared to simulated $\upalpha$-Ta powder diffraction. (b) Resistivity as a function of temperature for different growth temperatures. (c) Normalized resistance near the superconducting transition temperature of tantalum.}
    \label{fig:1}
    \makeatletter
    \let\save@currentlabel\@currentlabel
    \edef\@currentlabel{\save@currentlabel(a)}\label{fig:1a}
    \edef\@currentlabel{\save@currentlabel(b)}\label{fig:1b}
    \edef\@currentlabel{\save@currentlabel(c)}\label{fig:1c}
    \makeatother
\end{figure*}

\section{Results and Discussion}
Figure \ref{fig:1} shows the X-ray diffraction and electrical transport measurements of tantalum films grown at different manipulator temperatures on Al$_2$O$_3$(0001) substrates. In Figure \ref{fig:1a}, $\uptheta$-2$\uptheta$ scans were taken over a wide range to identify any possible orientations. For the 292 K grown film, a peak is observed near 33.5° corresponding to the diffraction of (002) planes in the $\upbeta$-Ta phase. In addition, two much broader peaks are observed; one labeled as $\upbeta$-Ta* and ranging from 37° to 42°, suggests the potential presence of (330), (202), (212), (411), (331) $\upbeta$ reflections \cite{lee2004texture} and another between 101° and 106°, denoted as $\upbeta$-Ta$^\dag$. The width of these peaks suggests a large angular distribution of grain orientation. 

On the other hand, the film grown at 7 K manipulator temperature displays multiple peaks within the 20 to 130-degree range shown in Figure \ref{fig:1a}. The most prominent peak is observed at 38.6°, closely aligning with the expected angle for $\upalpha$-Ta(110). Other peaks are found at 56.0°, 69.9°, 83.0°, 95.7°, 108.2°, and 122.3° corresponding to (200), (211), (220), (310), (222), (321) $\upalpha$-Ta reflections, respectively. The bottom of Figure  \ref{fig:1a} exhibits a simulated powder diffraction pattern for $\upalpha$-Ta. Remarkably, the peaks observed in the film grown at 7 K are close to the peaks found in the simulated $\upalpha$-Ta powder diffraction pattern with a lattice parameter of 3.3058 \AA. However, the measured lattice constants in all crystal directions are shifted compared to the simulation, potentially indicating an effect of thermal expansion mismatch. This similarity indicates the formation of a poly-crystalline $\upalpha$-Ta film with planes oriented similarly to $\upalpha$-Ta powder. There is no evidence for the presence of $\upbeta$-Ta in the XRD measurements in the film grown at 7K. 

\begin{figure}[b!]
    \centering
        \centering        \includegraphics{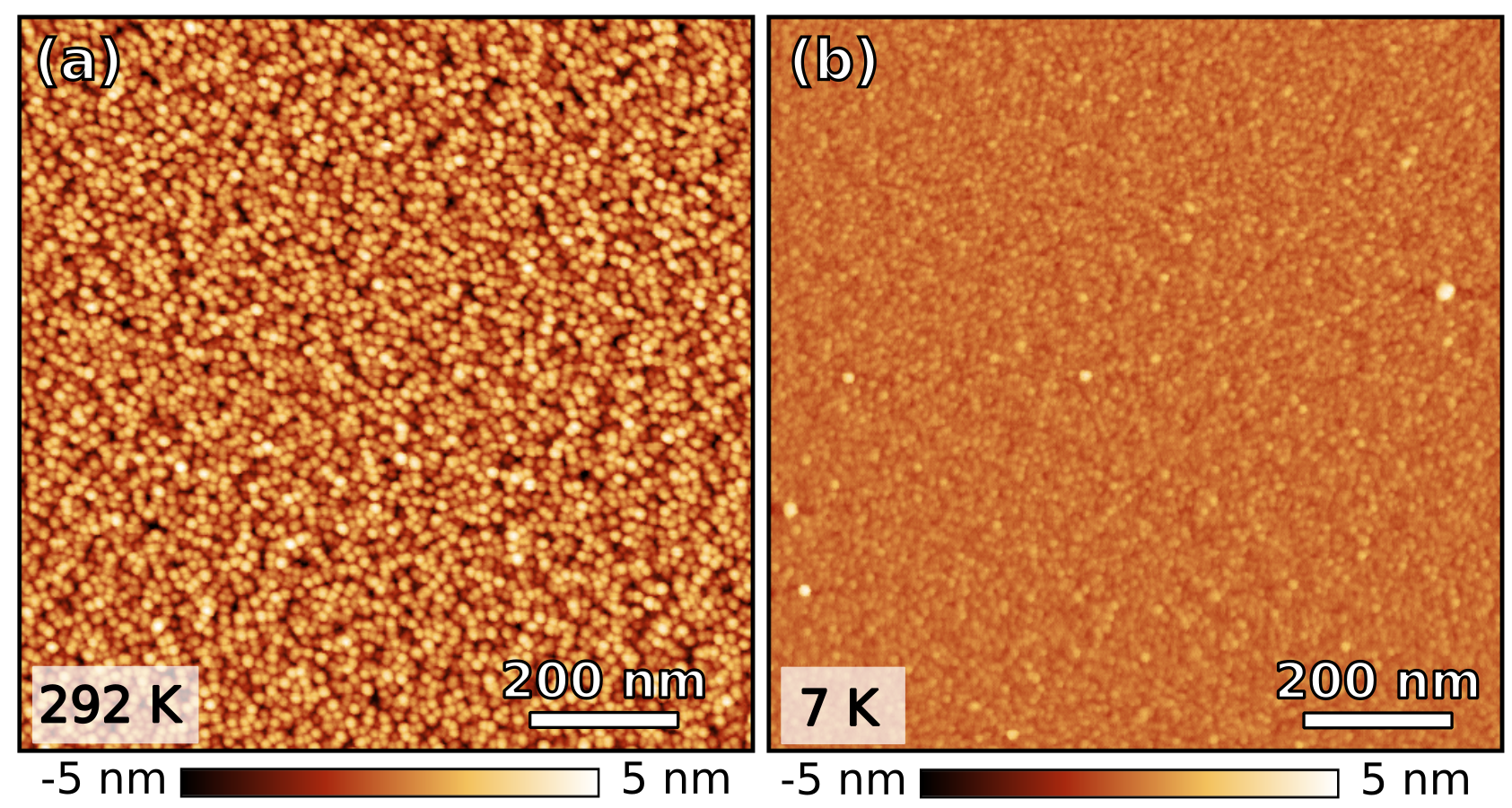} 
        \caption{Surface morphology of Ta grown on Al$_2$O$_3$(0001) at (a) 292 K and (b) 7 K by AFM.}
    \label{fig:2}
    \makeatletter
    \let\save@currentlabel\@currentlabel
    \edef\@currentlabel{\save@currentlabel(a)}\label{fig:2a}
    \edef\@currentlabel{\save@currentlabel(b)}\label{fig:2b}

    \makeatother
\end{figure}
\begin{figure*}[t!]
    \centering
        \centering        \includegraphics{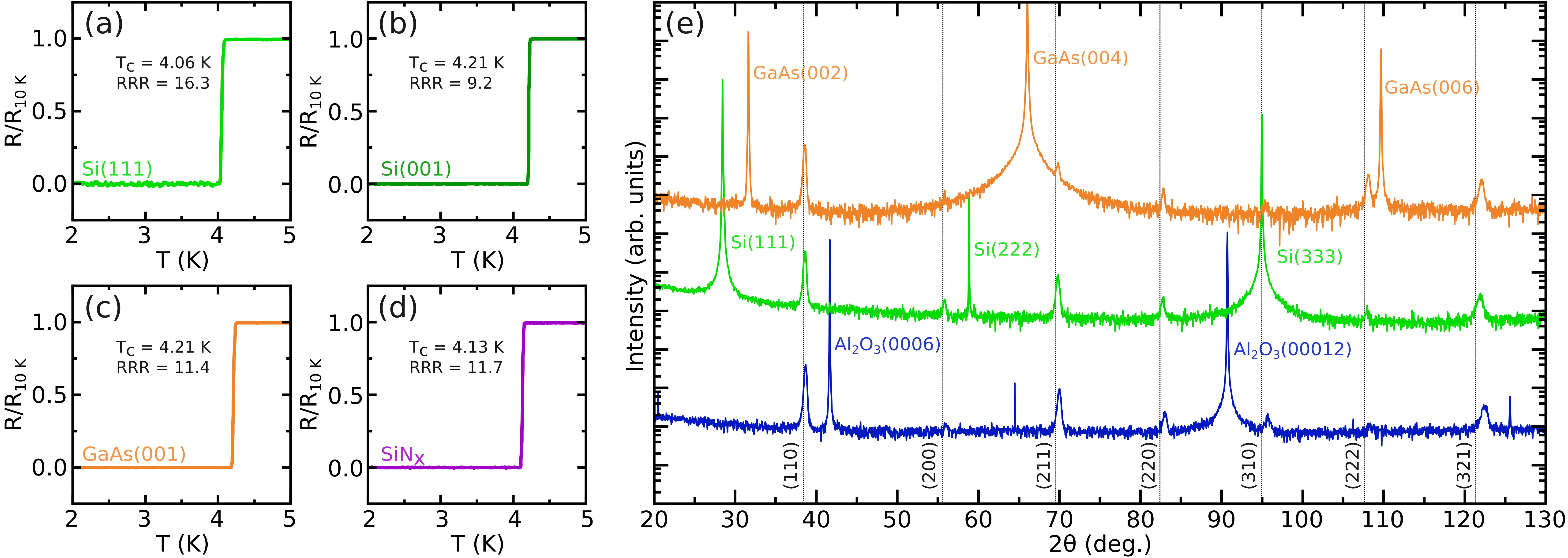} 
        \caption{Electrical and structural characterization of Tantalum grown at different substrates. (a) Normalized resistance of Ta grown on Si(111), (b) Si(001), (c) GaAs(001), (d) and SiN\textsubscript{x}. (e) XRD patterns of Ta on Al\textsubscript{2}O\textsubscript{3}(0001), Si(111), and GaAs(001).}
    \label{fig:3}
    \makeatletter
    \let\save@currentlabel\@currentlabel
    \edef\@currentlabel{\save@currentlabel(a)}\label{fig:3a}
    \edef\@currentlabel{\save@currentlabel(b)}\label{fig:3b}
    \edef\@currentlabel{\save@currentlabel(c)}\label{fig:3c}
    \edef\@currentlabel{\save@currentlabel(d)}\label{fig:3d}
    \edef\@currentlabel{\save@currentlabel(e)}\label{fig:3e}
    \makeatother
\end{figure*}
Examining Ta films grown at 50 K and 100 K, X-ray diffraction patterns reveal characteristics indicative of both $\upalpha$ and $\upbeta$  phases. For the film grown at 50 K, the primary observation is a significant decrease in the $\upalpha$-Ta (110) peak intensity, whereas the (211) peak stays relatively strong. Additionally, a weak peak near 33.5° indicates the presence of $\upbeta$-Ta. This suggests a predominant composition of $\upalpha$-Ta with a minor presence of $\upbeta$-Ta. The exact peak positions of $\upalpha$-Ta grown at 50 K are also shifted compared to the 7 K growth, providing evidence of the decreased effect of thermal expansion mismatch. Moving to the film grown at 100 K, a similar intensity of the $\upbeta$-Ta(002) is found compared to 292 K. Also the peak denoted as $\upbeta$-Ta* is present. The $\upbeta$-Ta$^\dag$ peak around 104° is not present. However, at 100 K there is a sharp peak on top of that corresponding to the presence of $\upalpha$-Ta (110), but other $\upalpha$-Ta peaks either decreased in intensity or disappeared. This shows that starting at low temperature the main phase is $\upalpha$-Ta, but as the growth temperature is increased, there will be an increased amount of $\upbeta$-Ta present. Finally, a $\upbeta$-Ta film is grown at room temperature with no evidence of $\upalpha$-Ta in the XRD measurement.

The temperature-dependent transport measurements in Figure \ref{fig:1b} and \ref{fig:1c} corroborate the X-ray diffraction results. Ta grown at 292 K exhibits a high resistivity of 181.8 $\upmu \Omega$\textit{·}cm at 300 K, in close agreement with the reported range of 160-180 $\upmu \Omega$\textit{·}cm for $\upbeta$-Ta \cite{catania1992low}. Moreover, this film does not show a superconducting transition above 2 K, the lowest temperature in the resistivity measurement system used for this study. The Residual-Resistance Ratio (RRR) of 1.05, defined as the ratio between resistivities at 300 K and 10 K, indicates substantial carrier scattering. These electrical characteristics are consistent with reported results for Ta thin films grown at room temperature \cite{read1965new, westwood1973effects, feinstein1973factors}. Upon reducing the growth temperature to 100 K, the film’s resistivity decreases to $\uprho$\textsubscript{300 K} = 70.3 $\upmu \Omega$\textit{·}cm, with an increased RRR of 4.6. Despite a drop in resistance below 3 K, superconductivity is not observed above 2 K. A further decrease in growth temperature results in a superconducting film with $T_C > 2$ K. For growth at 50 K, the film has a superconducting transition temperature, $T_C$ = 3.90 K and the film shows metallic behavior with $\uprho$\textsubscript{300 K} = 20.7 $\upmu \Omega$\textit{·}cm and a RRR = 7.2. The film with the most desirable superconducting properties is grown at 7 K, which shows a resistivity $\uprho$\textsubscript{300 K} of 13.4 µΩ·cm, very close to the bulk $\uprho$\textsubscript{300 K} = 13 $\upmu \Omega$\textit{·}cm for $\upalpha$-Ta \cite{catania1992low}. Furthermore, this film exhibits superconductivity with T\textsubscript{C} = 4.14 K, and a high RRR of 17.3.

\begin{figure}[b!]
    \centering
        \centering        \includegraphics{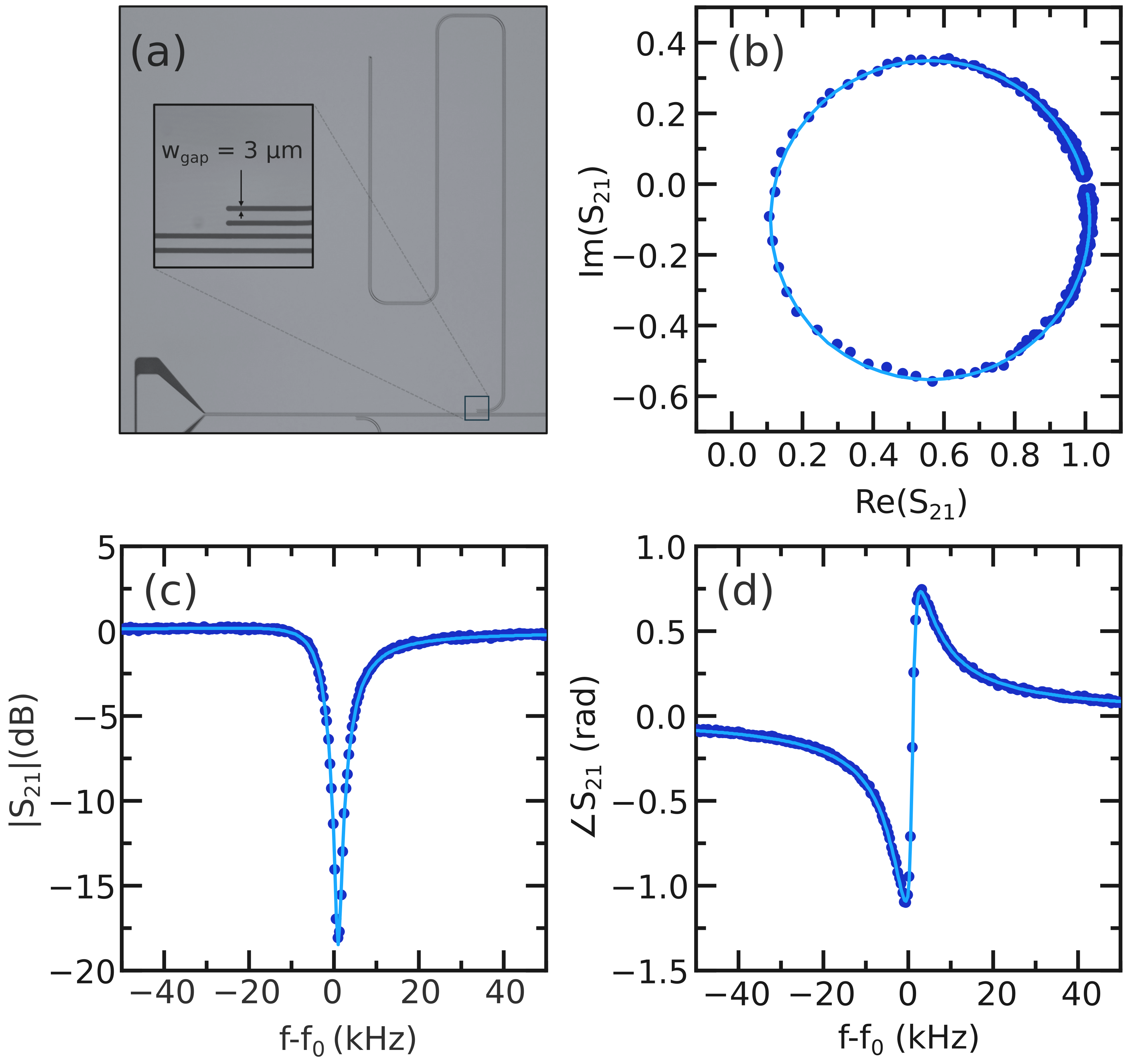} 
        \caption{Microwave characterization of Ta grown on Si(111). (a) Optical micrograph image of a CPW resonator with a gap width of 3 $\upmu$m. (b)-(d) Representative S\textsubscript{21} data along with the corresponding fits to the diameter correction method (DCM)}
    \label{fig:4}
    \makeatletter
    \let\save@currentlabel\@currentlabel
    \edef\@currentlabel{\save@currentlabel(a)}\label{fig:4a}
    \edef\@currentlabel{\save@currentlabel(b)}\label{fig:4b}
    \edef\@currentlabel{\save@currentlabel(c)}\label{fig:4c}
    \edef\@currentlabel{\save@currentlabel(d)}\label{fig:4d}
       \makeatother
\end{figure}
\begin{figure*}[t!]
    \centering
        \centering        \includegraphics{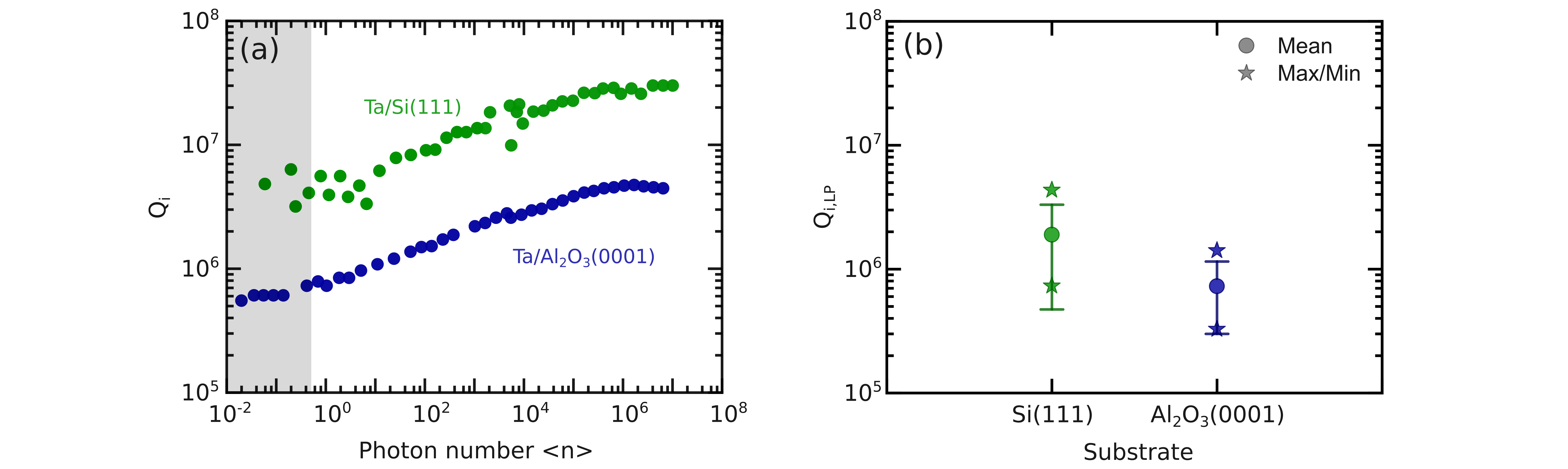} 
        \caption{Microwave characterization of Ta on Al$_2$O$_3$(0001) and Si(111). (a) Q\textsubscript{i} as a function of the applied microwave power, expressed in terms of average photon number in the cavity. (b) The average,  maximum, and minimum values for the Q\textsubscript{i,LP} plotted for the different substrates.}
    \label{fig:5}
    \makeatletter
    \let\save@currentlabel\@currentlabel
    \edef\@currentlabel{\save@currentlabel(a)}\label{fig:5a}
    \edef\@currentlabel{\save@currentlabel(b)}\label{fig:5b}
       \makeatother
\end{figure*}
To further examine the differences in films as a function of temperature, we conducted AFM measurements over a 1 $\upmu$m x 1 $\upmu$m area, as illustrated in Figure \ref{fig:2}. Here, the films grown at 7 K and 292 K are compared. Both films exhibit a granular structure characterized by grains that are uniformly distributed and similar in size. In the case of the film grown at 292 K, as shown in Figure \ref{fig:2a}, the root mean square (RMS) roughness is 1.57 nm. A notable reduction in RMS roughness is evident in the film grown at 7 K, as shown in Figure \ref{fig:2b}. In this case, the calculated RMS roughness is 0.45 nm, indicating that low temperature growth results in a significantly smoother surface. The typical surface morphology of $\upalpha$-Ta thin film grown on Al$_2$O$_3$(0001) consists of elongated and aligned grains that are indicative of textured $\upalpha$-Ta(110) films \cite{zhou2023epitaxial}. However, in the case of low temperature growth, none of these characteristic features in the grain structure are present. The absence of these features supports the formation of a polycrystalline film composed of randomly oriented $\upalpha$-Ta grains.

Next, the effect of substrate and substrate orientation on the low temperature growth of Ta is explored. Figure \ref{fig:3a}-\ref{fig:3d} shows the normalized resistance of Ta on Si(111), Si(001), GaAs(001), and SiNx. Remarkably, each of these Ta films grown on different surfaces exhibits a relatively sharp superconducting transition with T$_C$ ranging from 4.06 to 4.21 K. The RRR values have been determined to be 16.3, 9.2, 11.4, and 11.7 for the growth of Ta on Si(111), Si(001), GaAs(001), and SiN$_x$, respectively.  The XRD patterns of Ta on Si(111) and GaAs(001) are shown and compared to Ta on Al$_2$O$_3$(0001) in Figure \ref{fig:3e}. We observe the similar $\upalpha$-Ta peaks on Si(111) and GaAs(001), indicating that $\upalpha$-Ta grows in a polycrystalline structure independent of the substrate. 

CPW resonators are patterned using 50 nm Ta films grown at 7 K on both Si(111) and Al\textsubscript{2}O\textsubscript{3}(0001) to study and compare losses in these material systems. An optical microscope image of a resonator with a gap width of 3 $\upmu$m is shown in Figure \ref{fig:4a}.  Figure \ref{fig:4b}-\ref{fig:4d} shows representative S\textsubscript{21} data in hanger mode of the resonator for Ta on Si(111) along with the corresponding fits to the diameter correction method (DCM) \cite{khalil2012analysis} model which was used to extract internal and external quality factors Q\textsubscript{i} and Q\textsubscript{c}. A representative plot of Q\textsubscript{i} vs average photon number ($\langle n\rangle $) in the cavity for Ta on Si(111) and Ta on Al\textsubscript{2}O\textsubscript{3}(0001) is shown in Figure \ref{fig:5a}. The Q\textsubscript{i} vs $\langle n\rangle$ data is used to obtain the low power loss tangent ($\updelta_{LP}$), a commonly used figure of merit defined as 1/Q\textsubscript{i,LP}. Here Q\textsubscript{i,LP} is the low power quality factor taken at drive powers sufficiently low enough where Q\textsubscript{i} enhancement due to saturation of loss sources may be neglected. We take Q\textsubscript{i,LP} for each resonator to be an average of all ultra-low power data points below $\langle n\rangle $ = 0.5 photons. This low power threshold is indicated in Figure \ref{fig:5b}  and it may be seen that the Q\textsubscript{i} values trend towards saturation at ultra-low power. 

In Figure \ref{fig:5b}, the average,  maximum, and minimum values for the Q\textsubscript{i,LP} are plotted for the two different substrates. We found that the average Q\textsubscript{i,LP} = 1.9 million or $\updelta_{LP} = 5.3 \times 10^{-7}$ for Ta on Si(111), and Q\textsubscript{i,LP} = 729k or $\updelta_{LP} = 1.4 \times 10^{-6}$ for Ta on Al\textsubscript{2}O\textsubscript{3}(0001). In this study, the sample deposited on Si(111) substrate outperformed the samples deposited on sapphire. However, the fact that the Si sample was trenched while sapphire was not along with unknowns related to the loss associated with the substrate itself and its surface prohibits claiming one being inherently better than the other. Compared to the previous reports and taking into account the gap width of the resonator, an average loss of $\updelta_{LP} = 5.3 \times 10^{-7}$ is comparable to current state-of-the-art superconducting devices \cite{mcrae2020materials, jones2023grain}. 

The findings of our study regarding the growth and properties of cryogenically deposited Ta films prompt speculation on their crystallization behavior and its implications for superconducting circuit performance. We speculate that if the cryogenic deposited Ta film grows in a metastable amorphous state resulting from the low atomic mobility during deposition, it may recrystallize during warm-up into the thermodynamically stable $\upalpha$-Ta phase releasing latent heat to drive further crystallization. Our hypothesis aligns with the nucleation model proposed by Colin et al. \cite{colin2017origin}, which suggests direct nucleation of $\upalpha$- and $\upbeta$-Ta based on the growth temperature. However, in our scenario, the film does not form $\upbeta$-Ta, but instead, it adopts an amorphous state that lacks stability at room temperature. The recrystallization process within the amorphous film or on the surface explains the absence of an epitaxial relationship with the substrate, which leads to the formation of polycrystalline $\upalpha$-Ta, irrespective of the nature of the substrate. Further, epitaxial Ta may not be needed to get high-performing superconducting circuits contrary to what has been suggested in earlier work \cite{place2021new, wang2022towards, crowley2023disentangling}. Both our electrical transport and microwave measurements suggest that the polycrystalline $\upalpha$-Ta films are of high quality, and meet, or exceed the current state-of-the-art.

\section{Conclusion}
Ta films have been grown on c-plane sapphire at different cryogenic temperatures and room temperature. Our findings indicate that the electrical properties are optimal when Ta is grown at the lowest temperature possible. The structure of the films is identified as single phase $\upalpha$-Ta exhibiting a polycrystalline structure. An important advantage of this approach is the lack of epitaxy at low temperatures, allowing the $\upalpha$-Ta phase to be grown on either amorphous or crystalline substrates with comparable electrical properties. The microwave characterization of Ta on Si(111) and Ta on Al\textsubscript{2}O\textsubscript{3}(0001) shows high-performing resonators. Specifically, Si(111) demonstrated significantly better performance with a Q\textsubscript{i,LP} = 1.9 million compared to Al\textsubscript{2}O\textsubscript{3}(0001) with a with a Q\textsubscript{i,LP} = 0.7 million. This shows that low-loss superconducting circuits made from single phase polycrystalline $\upalpha$-Ta are promising despite the lack of epitaxy. 

The possibility of growing $\upalpha$-Ta on virtually any substrates opens up new avenues for integrating Ta into various applications that extend beyond superconducting quantum computing. For instance, in the field of topological quantum computing, superconductor-semiconductor heterostructures are a promising candidate \cite{lutchyn2018majorana}, with III-V materials like InAs and InSb commonly used. Generally, the choice of the superconductor material is limited due to conventional constraints regarding growth such as elevated growth temperatures on III-V materials. By using cryogenic growth, we successfully demonstrated the growth of single phase $\upalpha$-Ta on GaAs(001) with desirable superconducting properties.

\section{Acknowledgments}
We acknowledge the support of the New and Emerging Qubit Science and Technology (NEQST) Program initiated by the U.S. Army Research Office (ARO) under Grant No. W911NF2210052, and University of California, Santa Barbara (UCSB) National Science Foundation (NSF) Quantum Foundry through Q-AMASE-i Program via Award No. DMR-1906325. We also acknowledge the use of shared facilities of the UCSB MRSEC (NSF DMR–2308708) and the Nanotech UCSB Nanofabrication facility.  

\appendix 
\section{APPENDIX A: SURFACE PREPARATION}
The various substrates used in this study were prepared differently. 3” diameter Al\textsubscript{2}O\textsubscript{3}(0001) substrates were coated in resist and diced into 10 x 10 mm samples. The substrates were cleaned with solvents and etched in a 3:1 H\textsubscript{2}SO\textsubscript{4}:H\textsubscript{2}O\textsubscript{2} piranha solution for 5 minutes. Then, the substrates were loaded into the UHV system followed by a degas and anneal at a thermocouple temperature of 700°C. Silicon substrates were similarly diced and cleaned into 10 x 10 mm samples. The Si(111) substrates were intentionally oxidized by UV Ozone before loading into the UHV system. An \textit{in-situ} atomic hydrogen anneal at 800°C followed by a hydrogen-free anneal at 1000 °C resulted in a 7$\times$7 surface reconstruction observed by Reflection high-energy electron diffraction. Si(001) substrates were etched in 5\% HF right before loading in and annealed \textit{in-situ} at 600°C. GaAs used in this study consisted of an As\textsubscript{2} capped 500 nm GaAs unintentionally doped epilayer grown by MBE on a semi-insulating GaAs(001) substrate. The As cap was removed \textit{in-situ} by thermal desorption at 600 °C. The SiN\textsubscript{x} layer was deposited by Plasma Enhance Chemical Vapor Deposition on a Si(001) substrate. Then, the substrates were solvent-cleaned and annealed at 500°C to remove contaminants.

\bibliography{references.bib}

%apsrev4-2.bst 2019-01-14 (MD) hand-edited version of apsrev4-1.bst
%Control: key (0)
%Control: author (8) initials jnrlst
%Control: editor formatted (1) identically to author
%Control: production of article title (0) allowed
%Control: page (0) single
%Control: year (1) truncated
%Control: production of eprint (0) enabled
\begin{thebibliography}{42}%
\makeatletter
\providecommand \@ifxundefined [1]{%
 \@ifx{#1\undefined}
}%
\providecommand \@ifnum [1]{%
 \ifnum #1\expandafter \@firstoftwo
 \else \expandafter \@secondoftwo
 \fi
}%
\providecommand \@ifx [1]{%
 \ifx #1\expandafter \@firstoftwo
 \else \expandafter \@secondoftwo
 \fi
}%
\providecommand \natexlab [1]{#1}%
\providecommand \enquote  [1]{``#1''}%
\providecommand \bibnamefont  [1]{#1}%
\providecommand \bibfnamefont [1]{#1}%
\providecommand \citenamefont [1]{#1}%
\providecommand \href@noop [0]{\@secondoftwo}%
\providecommand \href [0]{\begingroup \@sanitize@url \@href}%
\providecommand \@href[1]{\@@startlink{#1}\@@href}%
\providecommand \@@href[1]{\endgroup#1\@@endlink}%
\providecommand \@sanitize@url [0]{\catcode `\\12\catcode `\$12\catcode `\&12\catcode `\#12\catcode `\^12\catcode `\_12\catcode `\%12\relax}%
\providecommand \@@startlink[1]{}%
\providecommand \@@endlink[0]{}%
\providecommand \url  [0]{\begingroup\@sanitize@url \@url }%
\providecommand \@url [1]{\endgroup\@href {#1}{\urlprefix }}%
\providecommand \urlprefix  [0]{URL }%
\providecommand \Eprint [0]{\href }%
\providecommand \doibase [0]{https://doi.org/}%
\providecommand \selectlanguage [0]{\@gobble}%
\providecommand \bibinfo  [0]{\@secondoftwo}%
\providecommand \bibfield  [0]{\@secondoftwo}%
\providecommand \translation [1]{[#1]}%
\providecommand \BibitemOpen [0]{}%
\providecommand \bibitemStop [0]{}%
\providecommand \bibitemNoStop [0]{.\EOS\space}%
\providecommand \EOS [0]{\spacefactor3000\relax}%
\providecommand \BibitemShut  [1]{\csname bibitem#1\endcsname}%
\let\auto@bib@innerbib\@empty
%</preamble>
\bibitem [{\citenamefont {Place}\ \emph {et~al.}(2021)\citenamefont {Place}, \citenamefont {Rodgers}, \citenamefont {Mundada}, \citenamefont {Smitham}, \citenamefont {Fitzpatrick}, \citenamefont {Leng}, \citenamefont {Premkumar}, \citenamefont {Bryon}, \citenamefont {Vrajitoarea}, \citenamefont {Sussman} \emph {et~al.}}]{place2021new}%
  \BibitemOpen
  \bibfield  {author} {\bibinfo {author} {\bibfnamefont {A.~P.~M.}\ \bibnamefont {Place}}, \bibinfo {author} {\bibfnamefont {L.~V.~H.}\ \bibnamefont {Rodgers}}, \bibinfo {author} {\bibfnamefont {P.}~\bibnamefont {Mundada}}, \bibinfo {author} {\bibfnamefont {B.~M.}\ \bibnamefont {Smitham}}, \bibinfo {author} {\bibfnamefont {M.}~\bibnamefont {Fitzpatrick}}, \bibinfo {author} {\bibfnamefont {Z.}~\bibnamefont {Leng}}, \bibinfo {author} {\bibfnamefont {A.}~\bibnamefont {Premkumar}}, \bibinfo {author} {\bibfnamefont {J.}~\bibnamefont {Bryon}}, \bibinfo {author} {\bibfnamefont {A.}~\bibnamefont {Vrajitoarea}}, \bibinfo {author} {\bibfnamefont {S.}~\bibnamefont {Sussman}}, \emph {et~al.},\ }\bibfield  {title} {\bibinfo {title} {New material platform for superconducting transmon qubits with coherence times exceeding 0.3 milliseconds},\ }\href@noop {} {\bibfield  {journal} {\bibinfo  {journal} {Nature communications}\ }\textbf {\bibinfo {volume} {12}},\ \bibinfo {pages} {1779} (\bibinfo {year} {2021})}\BibitemShut
  {NoStop}%
\bibitem [{\citenamefont {Wang}\ \emph {et~al.}(2022)\citenamefont {Wang}, \citenamefont {Li}, \citenamefont {Xu}, \citenamefont {Li}, \citenamefont {Wang}, \citenamefont {Yang}, \citenamefont {Mi}, \citenamefont {Liang}, \citenamefont {Su}, \citenamefont {Yang} \emph {et~al.}}]{wang2022towards}%
  \BibitemOpen
  \bibfield  {author} {\bibinfo {author} {\bibfnamefont {C.}~\bibnamefont {Wang}}, \bibinfo {author} {\bibfnamefont {X.}~\bibnamefont {Li}}, \bibinfo {author} {\bibfnamefont {H.}~\bibnamefont {Xu}}, \bibinfo {author} {\bibfnamefont {Z.}~\bibnamefont {Li}}, \bibinfo {author} {\bibfnamefont {J.}~\bibnamefont {Wang}}, \bibinfo {author} {\bibfnamefont {Z.}~\bibnamefont {Yang}}, \bibinfo {author} {\bibfnamefont {Z.}~\bibnamefont {Mi}}, \bibinfo {author} {\bibfnamefont {X.}~\bibnamefont {Liang}}, \bibinfo {author} {\bibfnamefont {T.}~\bibnamefont {Su}}, \bibinfo {author} {\bibfnamefont {C.}~\bibnamefont {Yang}}, \emph {et~al.},\ }\bibfield  {title} {\bibinfo {title} {Towards practical quantum computers: Transmon qubit with a lifetime approaching 0.5 milliseconds},\ }\href@noop {} {\bibfield  {journal} {\bibinfo  {journal} {npj Quantum Information}\ }\textbf {\bibinfo {volume} {8}},\ \bibinfo {pages} {3} (\bibinfo {year} {2022})}\BibitemShut {NoStop}%
\bibitem [{\citenamefont {Wang}\ \emph {et~al.}(2015)\citenamefont {Wang}, \citenamefont {Axline}, \citenamefont {Gao}, \citenamefont {Brecht}, \citenamefont {Chu}, \citenamefont {Frunzio}, \citenamefont {Devoret},\ and\ \citenamefont {Schoelkopf}}]{wang2015surface}%
  \BibitemOpen
  \bibfield  {author} {\bibinfo {author} {\bibfnamefont {C.}~\bibnamefont {Wang}}, \bibinfo {author} {\bibfnamefont {C.}~\bibnamefont {Axline}}, \bibinfo {author} {\bibfnamefont {Y.~Y.}\ \bibnamefont {Gao}}, \bibinfo {author} {\bibfnamefont {T.}~\bibnamefont {Brecht}}, \bibinfo {author} {\bibfnamefont {Y.}~\bibnamefont {Chu}}, \bibinfo {author} {\bibfnamefont {L.}~\bibnamefont {Frunzio}}, \bibinfo {author} {\bibfnamefont {M.~H.}\ \bibnamefont {Devoret}},\ and\ \bibinfo {author} {\bibfnamefont {R.~J.}\ \bibnamefont {Schoelkopf}},\ }\bibfield  {title} {\bibinfo {title} {Surface participation and dielectric loss in superconducting qubits},\ }\href@noop {} {\bibfield  {journal} {\bibinfo  {journal} {Applied Physics Letters}\ }\textbf {\bibinfo {volume} {107}} (\bibinfo {year} {2015})}\BibitemShut {NoStop}%
\bibitem [{\citenamefont {Dial}\ \emph {et~al.}(2016)\citenamefont {Dial}, \citenamefont {McClure}, \citenamefont {Poletto}, \citenamefont {Keefe}, \citenamefont {Rothwell}, \citenamefont {Gambetta}, \citenamefont {Abraham}, \citenamefont {Chow},\ and\ \citenamefont {Steffen}}]{dial2016bulk}%
  \BibitemOpen
  \bibfield  {author} {\bibinfo {author} {\bibfnamefont {O.}~\bibnamefont {Dial}}, \bibinfo {author} {\bibfnamefont {D.~T.}\ \bibnamefont {McClure}}, \bibinfo {author} {\bibfnamefont {S.}~\bibnamefont {Poletto}}, \bibinfo {author} {\bibfnamefont {G.~A.}\ \bibnamefont {Keefe}}, \bibinfo {author} {\bibfnamefont {M.~B.}\ \bibnamefont {Rothwell}}, \bibinfo {author} {\bibfnamefont {J.~M.}\ \bibnamefont {Gambetta}}, \bibinfo {author} {\bibfnamefont {D.~W.}\ \bibnamefont {Abraham}}, \bibinfo {author} {\bibfnamefont {J.~M.}\ \bibnamefont {Chow}},\ and\ \bibinfo {author} {\bibfnamefont {M.}~\bibnamefont {Steffen}},\ }\bibfield  {title} {\bibinfo {title} {Bulk and surface loss in superconducting transmon qubits},\ }\href@noop {} {\bibfield  {journal} {\bibinfo  {journal} {Superconductor Science and Technology}\ }\textbf {\bibinfo {volume} {29}},\ \bibinfo {pages} {044001} (\bibinfo {year} {2016})}\BibitemShut {NoStop}%
\bibitem [{\citenamefont {Gambetta}\ \emph {et~al.}(2016)\citenamefont {Gambetta}, \citenamefont {Murray}, \citenamefont {Fung}, \citenamefont {McClure}, \citenamefont {Dial}, \citenamefont {Shanks}, \citenamefont {Sleight},\ and\ \citenamefont {Steffen}}]{gambetta2016investigating}%
  \BibitemOpen
  \bibfield  {author} {\bibinfo {author} {\bibfnamefont {J.~M.}\ \bibnamefont {Gambetta}}, \bibinfo {author} {\bibfnamefont {C.~E.}\ \bibnamefont {Murray}}, \bibinfo {author} {\bibfnamefont {Y.-K.-K.}\ \bibnamefont {Fung}}, \bibinfo {author} {\bibfnamefont {D.~T.}\ \bibnamefont {McClure}}, \bibinfo {author} {\bibfnamefont {O.}~\bibnamefont {Dial}}, \bibinfo {author} {\bibfnamefont {W.}~\bibnamefont {Shanks}}, \bibinfo {author} {\bibfnamefont {J.~W.}\ \bibnamefont {Sleight}},\ and\ \bibinfo {author} {\bibfnamefont {M.}~\bibnamefont {Steffen}},\ }\bibfield  {title} {\bibinfo {title} {Investigating surface loss effects in superconducting transmon qubits},\ }\href@noop {} {\bibfield  {journal} {\bibinfo  {journal} {IEEE Transactions on Applied Superconductivity}\ }\textbf {\bibinfo {volume} {27}},\ \bibinfo {pages} {1} (\bibinfo {year} {2016})}\BibitemShut {NoStop}%
\bibitem [{\citenamefont {Martinis}\ \emph {et~al.}(2005)\citenamefont {Martinis}, \citenamefont {Cooper}, \citenamefont {McDermott}, \citenamefont {Steffen}, \citenamefont {Ansmann}, \citenamefont {Osborn}, \citenamefont {Cicak}, \citenamefont {Oh}, \citenamefont {Pappas}, \citenamefont {Simmonds} \emph {et~al.}}]{martinis2005decoherence}%
  \BibitemOpen
  \bibfield  {author} {\bibinfo {author} {\bibfnamefont {J.~M.}\ \bibnamefont {Martinis}}, \bibinfo {author} {\bibfnamefont {K.~B.}\ \bibnamefont {Cooper}}, \bibinfo {author} {\bibfnamefont {R.}~\bibnamefont {McDermott}}, \bibinfo {author} {\bibfnamefont {M.}~\bibnamefont {Steffen}}, \bibinfo {author} {\bibfnamefont {M.}~\bibnamefont {Ansmann}}, \bibinfo {author} {\bibfnamefont {K.~D.}\ \bibnamefont {Osborn}}, \bibinfo {author} {\bibfnamefont {K.}~\bibnamefont {Cicak}}, \bibinfo {author} {\bibfnamefont {S.}~\bibnamefont {Oh}}, \bibinfo {author} {\bibfnamefont {D.~P.}\ \bibnamefont {Pappas}}, \bibinfo {author} {\bibfnamefont {R.~W.}\ \bibnamefont {Simmonds}}, \emph {et~al.},\ }\bibfield  {title} {\bibinfo {title} {Decoherence in josephson qubits from dielectric loss},\ }\href@noop {} {\bibfield  {journal} {\bibinfo  {journal} {Physical review letters}\ }\textbf {\bibinfo {volume} {95}},\ \bibinfo {pages} {210503} (\bibinfo {year} {2005})}\BibitemShut {NoStop}%
\bibitem [{\citenamefont {Pappas}\ \emph {et~al.}(2011)\citenamefont {Pappas}, \citenamefont {Vissers}, \citenamefont {Wisbey}, \citenamefont {Kline},\ and\ \citenamefont {Gao}}]{pappas2011two}%
  \BibitemOpen
  \bibfield  {author} {\bibinfo {author} {\bibfnamefont {D.~P.}\ \bibnamefont {Pappas}}, \bibinfo {author} {\bibfnamefont {M.~R.}\ \bibnamefont {Vissers}}, \bibinfo {author} {\bibfnamefont {D.~S.}\ \bibnamefont {Wisbey}}, \bibinfo {author} {\bibfnamefont {J.~S.}\ \bibnamefont {Kline}},\ and\ \bibinfo {author} {\bibfnamefont {J.}~\bibnamefont {Gao}},\ }\bibfield  {title} {\bibinfo {title} {Two level system loss in superconducting microwave resonators},\ }\href@noop {} {\bibfield  {journal} {\bibinfo  {journal} {IEEE Transactions on Applied Superconductivity}\ }\textbf {\bibinfo {volume} {21}},\ \bibinfo {pages} {871} (\bibinfo {year} {2011})}\BibitemShut {NoStop}%
\bibitem [{\citenamefont {Lisenfeld}\ \emph {et~al.}(2019)\citenamefont {Lisenfeld}, \citenamefont {Bilmes}, \citenamefont {Megrant}, \citenamefont {Barends}, \citenamefont {Kelly}, \citenamefont {Klimov}, \citenamefont {Weiss}, \citenamefont {Martinis},\ and\ \citenamefont {Ustinov}}]{lisenfeld2019electric}%
  \BibitemOpen
  \bibfield  {author} {\bibinfo {author} {\bibfnamefont {J.}~\bibnamefont {Lisenfeld}}, \bibinfo {author} {\bibfnamefont {A.}~\bibnamefont {Bilmes}}, \bibinfo {author} {\bibfnamefont {A.}~\bibnamefont {Megrant}}, \bibinfo {author} {\bibfnamefont {R.}~\bibnamefont {Barends}}, \bibinfo {author} {\bibfnamefont {J.}~\bibnamefont {Kelly}}, \bibinfo {author} {\bibfnamefont {P.}~\bibnamefont {Klimov}}, \bibinfo {author} {\bibfnamefont {G.}~\bibnamefont {Weiss}}, \bibinfo {author} {\bibfnamefont {J.~M.}\ \bibnamefont {Martinis}},\ and\ \bibinfo {author} {\bibfnamefont {A.~V.}\ \bibnamefont {Ustinov}},\ }\bibfield  {title} {\bibinfo {title} {Electric field spectroscopy of material defects in transmon qubits},\ }\href@noop {} {\bibfield  {journal} {\bibinfo  {journal} {npj Quantum Information}\ }\textbf {\bibinfo {volume} {5}},\ \bibinfo {pages} {105} (\bibinfo {year} {2019})}\BibitemShut {NoStop}%
\bibitem [{\citenamefont {Wang}\ \emph {et~al.}(2014)\citenamefont {Wang}, \citenamefont {Gao}, \citenamefont {Pop}, \citenamefont {Vool}, \citenamefont {Axline}, \citenamefont {Brecht}, \citenamefont {Heeres}, \citenamefont {Frunzio}, \citenamefont {Devoret}, \citenamefont {Catelani} \emph {et~al.}}]{wang2014measurement}%
  \BibitemOpen
  \bibfield  {author} {\bibinfo {author} {\bibfnamefont {C.}~\bibnamefont {Wang}}, \bibinfo {author} {\bibfnamefont {Y.~Y.}\ \bibnamefont {Gao}}, \bibinfo {author} {\bibfnamefont {I.~M.}\ \bibnamefont {Pop}}, \bibinfo {author} {\bibfnamefont {U.}~\bibnamefont {Vool}}, \bibinfo {author} {\bibfnamefont {C.}~\bibnamefont {Axline}}, \bibinfo {author} {\bibfnamefont {T.}~\bibnamefont {Brecht}}, \bibinfo {author} {\bibfnamefont {R.~W.}\ \bibnamefont {Heeres}}, \bibinfo {author} {\bibfnamefont {L.}~\bibnamefont {Frunzio}}, \bibinfo {author} {\bibfnamefont {M.~H.}\ \bibnamefont {Devoret}}, \bibinfo {author} {\bibfnamefont {G.}~\bibnamefont {Catelani}}, \emph {et~al.},\ }\bibfield  {title} {\bibinfo {title} {Measurement and control of quasiparticle dynamics in a superconducting qubit},\ }\href@noop {} {\bibfield  {journal} {\bibinfo  {journal} {Nature communications}\ }\textbf {\bibinfo {volume} {5}},\ \bibinfo {pages} {5836} (\bibinfo {year} {2014})}\BibitemShut {NoStop}%
\bibitem [{\citenamefont {Mazin}(2022)}]{mazin2022superconducting}%
  \BibitemOpen
  \bibfield  {author} {\bibinfo {author} {\bibfnamefont {B.~A.}\ \bibnamefont {Mazin}},\ }\bibfield  {title} {\bibinfo {title} {Superconducting materials for microwave kinetic inductance detectors},\ }in\ \href@noop {} {\emph {\bibinfo {booktitle} {Handbook of Superconductivity}}}\ (\bibinfo  {publisher} {CRC Press},\ \bibinfo {year} {2022})\ pp.\ \bibinfo {pages} {756--765}\BibitemShut {NoStop}%
\bibitem [{\citenamefont {Crowley}\ \emph {et~al.}(2023)\citenamefont {Crowley}, \citenamefont {McLellan}, \citenamefont {Dutta}, \citenamefont {Shumiya}, \citenamefont {Place}, \citenamefont {Le}, \citenamefont {Gang}, \citenamefont {Madhavan}, \citenamefont {Bland}, \citenamefont {Chang} \emph {et~al.}}]{crowley2023disentangling}%
  \BibitemOpen
  \bibfield  {author} {\bibinfo {author} {\bibfnamefont {K.~D.}\ \bibnamefont {Crowley}}, \bibinfo {author} {\bibfnamefont {R.~A.}\ \bibnamefont {McLellan}}, \bibinfo {author} {\bibfnamefont {A.}~\bibnamefont {Dutta}}, \bibinfo {author} {\bibfnamefont {N.}~\bibnamefont {Shumiya}}, \bibinfo {author} {\bibfnamefont {A.~P.~M.}\ \bibnamefont {Place}}, \bibinfo {author} {\bibfnamefont {X.~H.}\ \bibnamefont {Le}}, \bibinfo {author} {\bibfnamefont {Y.}~\bibnamefont {Gang}}, \bibinfo {author} {\bibfnamefont {T.}~\bibnamefont {Madhavan}}, \bibinfo {author} {\bibfnamefont {M.~P.}\ \bibnamefont {Bland}}, \bibinfo {author} {\bibfnamefont {R.}~\bibnamefont {Chang}}, \emph {et~al.},\ }\bibfield  {title} {\bibinfo {title} {Disentangling losses in tantalum superconducting circuits},\ }\href@noop {} {\bibfield  {journal} {\bibinfo  {journal} {Physical Review X}\ }\textbf {\bibinfo {volume} {13}},\ \bibinfo {pages} {041005} (\bibinfo {year} {2023})}\BibitemShut {NoStop}%
\bibitem [{\citenamefont {McLellan}\ \emph {et~al.}(2023)\citenamefont {McLellan}, \citenamefont {Dutta}, \citenamefont {Zhou}, \citenamefont {Jia}, \citenamefont {Weiland}, \citenamefont {Gui}, \citenamefont {Place}, \citenamefont {Crowley}, \citenamefont {Le}, \citenamefont {Madhavan} \emph {et~al.}}]{mclellan2023chemical}%
  \BibitemOpen
  \bibfield  {author} {\bibinfo {author} {\bibfnamefont {R.~A.}\ \bibnamefont {McLellan}}, \bibinfo {author} {\bibfnamefont {A.}~\bibnamefont {Dutta}}, \bibinfo {author} {\bibfnamefont {C.}~\bibnamefont {Zhou}}, \bibinfo {author} {\bibfnamefont {Y.}~\bibnamefont {Jia}}, \bibinfo {author} {\bibfnamefont {C.}~\bibnamefont {Weiland}}, \bibinfo {author} {\bibfnamefont {X.}~\bibnamefont {Gui}}, \bibinfo {author} {\bibfnamefont {A.~P.~M.}\ \bibnamefont {Place}}, \bibinfo {author} {\bibfnamefont {K.~D.}\ \bibnamefont {Crowley}}, \bibinfo {author} {\bibfnamefont {X.~H.}\ \bibnamefont {Le}}, \bibinfo {author} {\bibfnamefont {T.}~\bibnamefont {Madhavan}}, \emph {et~al.},\ }\bibfield  {title} {\bibinfo {title} {Chemical profiles of the oxides on tantalum in state of the art superconducting circuits},\ }\href@noop {} {\bibfield  {journal} {\bibinfo  {journal} {Advanced Science}\ }\textbf {\bibinfo {volume} {10}},\ \bibinfo {pages} {2300921} (\bibinfo {year} {2023})}\BibitemShut {NoStop}%
\bibitem [{\citenamefont {Read}\ and\ \citenamefont {Altman}(1965)}]{read1965new}%
  \BibitemOpen
  \bibfield  {author} {\bibinfo {author} {\bibfnamefont {M.~H.}\ \bibnamefont {Read}}\ and\ \bibinfo {author} {\bibfnamefont {C.}~\bibnamefont {Altman}},\ }\bibfield  {title} {\bibinfo {title} {A new structure in tantalum thin films},\ }\href@noop {} {\bibfield  {journal} {\bibinfo  {journal} {Applied Physics Letters}\ }\textbf {\bibinfo {volume} {7}},\ \bibinfo {pages} {51} (\bibinfo {year} {1965})}\BibitemShut {NoStop}%
\bibitem [{\citenamefont {Westwood}\ \emph {et~al.}(1973)\citenamefont {Westwood}, \citenamefont {Boynton},\ and\ \citenamefont {Wilcox}}]{westwood1973effects}%
  \BibitemOpen
  \bibfield  {author} {\bibinfo {author} {\bibfnamefont {W.~D.}\ \bibnamefont {Westwood}}, \bibinfo {author} {\bibfnamefont {R.~J.}\ \bibnamefont {Boynton}},\ and\ \bibinfo {author} {\bibfnamefont {P.~S.}\ \bibnamefont {Wilcox}},\ }\bibfield  {title} {\bibinfo {title} {The effects of argon pressure on the properties of sputtered tantalum films},\ }\href@noop {} {\bibfield  {journal} {\bibinfo  {journal} {Thin Solid Films}\ }\textbf {\bibinfo {volume} {16}},\ \bibinfo {pages} {1} (\bibinfo {year} {1973})}\BibitemShut {NoStop}%
\bibitem [{\citenamefont {Feinstein}\ and\ \citenamefont {Huttemann}(1973)}]{feinstein1973factors}%
  \BibitemOpen
  \bibfield  {author} {\bibinfo {author} {\bibfnamefont {L.~G.}\ \bibnamefont {Feinstein}}\ and\ \bibinfo {author} {\bibfnamefont {R.~D.}\ \bibnamefont {Huttemann}},\ }\bibfield  {title} {\bibinfo {title} {Factors controlling the structure of sputtered ta films},\ }\href@noop {} {\bibfield  {journal} {\bibinfo  {journal} {Thin Solid Films}\ }\textbf {\bibinfo {volume} {16}},\ \bibinfo {pages} {129} (\bibinfo {year} {1973})}\BibitemShut {NoStop}%
\bibitem [{\citenamefont {Colin}\ \emph {et~al.}(2017)\citenamefont {Colin}, \citenamefont {Abadias}, \citenamefont {Michel},\ and\ \citenamefont {Jaouen}}]{colin2017origin}%
  \BibitemOpen
  \bibfield  {author} {\bibinfo {author} {\bibfnamefont {J.~J.}\ \bibnamefont {Colin}}, \bibinfo {author} {\bibfnamefont {G.}~\bibnamefont {Abadias}}, \bibinfo {author} {\bibfnamefont {A.}~\bibnamefont {Michel}},\ and\ \bibinfo {author} {\bibfnamefont {C.}~\bibnamefont {Jaouen}},\ }\bibfield  {title} {\bibinfo {title} {On the origin of the metastable $\beta$-ta phase stabilization in tantalum sputtered thin films},\ }\href@noop {} {\bibfield  {journal} {\bibinfo  {journal} {Acta Materialia}\ }\textbf {\bibinfo {volume} {126}},\ \bibinfo {pages} {481} (\bibinfo {year} {2017})}\BibitemShut {NoStop}%
\bibitem [{\citenamefont {Matson}\ \emph {et~al.}(2000)\citenamefont {Matson}, \citenamefont {McClanahan}, \citenamefont {Rice}, \citenamefont {Lee},\ and\ \citenamefont {Windover}}]{matson2000effect}%
  \BibitemOpen
  \bibfield  {author} {\bibinfo {author} {\bibfnamefont {D.~W.}\ \bibnamefont {Matson}}, \bibinfo {author} {\bibfnamefont {E.~D.}\ \bibnamefont {McClanahan}}, \bibinfo {author} {\bibfnamefont {J.~P.}\ \bibnamefont {Rice}}, \bibinfo {author} {\bibfnamefont {S.~L.}\ \bibnamefont {Lee}},\ and\ \bibinfo {author} {\bibfnamefont {D.}~\bibnamefont {Windover}},\ }\bibfield  {title} {\bibinfo {title} {Effect of sputtering parameters on ta coatings for gun bore applications},\ }\href@noop {} {\bibfield  {journal} {\bibinfo  {journal} {Surface and Coatings Technology}\ }\textbf {\bibinfo {volume} {133}},\ \bibinfo {pages} {411} (\bibinfo {year} {2000})}\BibitemShut {NoStop}%
\bibitem [{\citenamefont {Gladczuk}\ \emph {et~al.}(2004)\citenamefont {Gladczuk}, \citenamefont {Patel}, \citenamefont {Paur},\ and\ \citenamefont {Sosnowski}}]{gladczuk2004tantalum}%
  \BibitemOpen
  \bibfield  {author} {\bibinfo {author} {\bibfnamefont {L.}~\bibnamefont {Gladczuk}}, \bibinfo {author} {\bibfnamefont {A.}~\bibnamefont {Patel}}, \bibinfo {author} {\bibfnamefont {C.~S.}\ \bibnamefont {Paur}},\ and\ \bibinfo {author} {\bibfnamefont {M.}~\bibnamefont {Sosnowski}},\ }\bibfield  {title} {\bibinfo {title} {Tantalum films for protective coatings of steel},\ }\href@noop {} {\bibfield  {journal} {\bibinfo  {journal} {Thin Solid Films}\ }\textbf {\bibinfo {volume} {467}},\ \bibinfo {pages} {150} (\bibinfo {year} {2004})}\BibitemShut {NoStop}%
\bibitem [{\citenamefont {Zhou}\ \emph {et~al.}(2009)\citenamefont {Zhou}, \citenamefont {Xie}, \citenamefont {Xiao}, \citenamefont {Hu},\ and\ \citenamefont {He}}]{zhou2009effects}%
  \BibitemOpen
  \bibfield  {author} {\bibinfo {author} {\bibfnamefont {Y.~M.}\ \bibnamefont {Zhou}}, \bibinfo {author} {\bibfnamefont {Z.}~\bibnamefont {Xie}}, \bibinfo {author} {\bibfnamefont {H.~N.}\ \bibnamefont {Xiao}}, \bibinfo {author} {\bibfnamefont {P.~F.}\ \bibnamefont {Hu}},\ and\ \bibinfo {author} {\bibfnamefont {J.}~\bibnamefont {He}},\ }\bibfield  {title} {\bibinfo {title} {Effects of deposition parameters on tantalum films deposited by direct current magnetron sputtering},\ }\href@noop {} {\bibfield  {journal} {\bibinfo  {journal} {Journal of Vacuum Science \& Technology A: Vacuum, Surfaces, and Films}\ }\textbf {\bibinfo {volume} {27}},\ \bibinfo {pages} {109} (\bibinfo {year} {2009})}\BibitemShut {NoStop}%
\bibitem [{\citenamefont {Zhou}\ \emph {et~al.}(2023)\citenamefont {Zhou}, \citenamefont {Yang}, \citenamefont {Wang}, \citenamefont {Wu}, \citenamefont {Xiong},\ and\ \citenamefont {Feng}}]{zhou2023epitaxial}%
  \BibitemOpen
  \bibfield  {author} {\bibinfo {author} {\bibfnamefont {B.}~\bibnamefont {Zhou}}, \bibinfo {author} {\bibfnamefont {L.}~\bibnamefont {Yang}}, \bibinfo {author} {\bibfnamefont {T.}~\bibnamefont {Wang}}, \bibinfo {author} {\bibfnamefont {Y.}~\bibnamefont {Wu}}, \bibinfo {author} {\bibfnamefont {K.}~\bibnamefont {Xiong}},\ and\ \bibinfo {author} {\bibfnamefont {J.}~\bibnamefont {Feng}},\ }\bibfield  {title} {\bibinfo {title} {Epitaxial alpha-ta (110) film on a-plane sapphire substrate for superconducting qubits with long coherence times},\ }\href@noop {} {\bibfield  {journal} {\bibinfo  {journal} {arXiv preprint arXiv:2306.09568}\ } (\bibinfo {year} {2023})}\BibitemShut {NoStop}%
\bibitem [{\citenamefont {Shi}\ \emph {et~al.}(2022)\citenamefont {Shi}, \citenamefont {Guo}, \citenamefont {Su}, \citenamefont {Chi}, \citenamefont {Sheng}, \citenamefont {Jiang}, \citenamefont {Cao}, \citenamefont {Wu}, \citenamefont {Tu}, \citenamefont {Sun} \emph {et~al.}}]{shi2022tantalum}%
  \BibitemOpen
  \bibfield  {author} {\bibinfo {author} {\bibfnamefont {L.}~\bibnamefont {Shi}}, \bibinfo {author} {\bibfnamefont {T.}~\bibnamefont {Guo}}, \bibinfo {author} {\bibfnamefont {R.}~\bibnamefont {Su}}, \bibinfo {author} {\bibfnamefont {T.}~\bibnamefont {Chi}}, \bibinfo {author} {\bibfnamefont {Y.}~\bibnamefont {Sheng}}, \bibinfo {author} {\bibfnamefont {J.}~\bibnamefont {Jiang}}, \bibinfo {author} {\bibfnamefont {C.}~\bibnamefont {Cao}}, \bibinfo {author} {\bibfnamefont {J.}~\bibnamefont {Wu}}, \bibinfo {author} {\bibfnamefont {X.}~\bibnamefont {Tu}}, \bibinfo {author} {\bibfnamefont {G.}~\bibnamefont {Sun}}, \emph {et~al.},\ }\bibfield  {title} {\bibinfo {title} {Tantalum microwave resonators with ultra-high intrinsic quality factors},\ }\href@noop {} {\bibfield  {journal} {\bibinfo  {journal} {Applied Physics Letters}\ }\textbf {\bibinfo {volume} {121}} (\bibinfo {year} {2022})}\BibitemShut {NoStop}%
\bibitem [{\citenamefont {Lozano}\ \emph {et~al.}(2022)\citenamefont {Lozano}, \citenamefont {Mongillo}, \citenamefont {Piao}, \citenamefont {Couet}, \citenamefont {Wan}, \citenamefont {Canvel}, \citenamefont {Vadiraj}, \citenamefont {Ivanov}, \citenamefont {Verjauw}, \citenamefont {Acharya} \emph {et~al.}}]{lozano2022manufacturing}%
  \BibitemOpen
  \bibfield  {author} {\bibinfo {author} {\bibfnamefont {D.~P.}\ \bibnamefont {Lozano}}, \bibinfo {author} {\bibfnamefont {M.}~\bibnamefont {Mongillo}}, \bibinfo {author} {\bibfnamefont {X.}~\bibnamefont {Piao}}, \bibinfo {author} {\bibfnamefont {S.}~\bibnamefont {Couet}}, \bibinfo {author} {\bibfnamefont {D.}~\bibnamefont {Wan}}, \bibinfo {author} {\bibfnamefont {Y.}~\bibnamefont {Canvel}}, \bibinfo {author} {\bibfnamefont {A.~M.}\ \bibnamefont {Vadiraj}}, \bibinfo {author} {\bibfnamefont {T.}~\bibnamefont {Ivanov}}, \bibinfo {author} {\bibfnamefont {J.}~\bibnamefont {Verjauw}}, \bibinfo {author} {\bibfnamefont {R.}~\bibnamefont {Acharya}}, \emph {et~al.},\ }\bibfield  {title} {\bibinfo {title} {Manufacturing high-q superconducting alpha - tantalum resonators on silicon wafers},\ }\href@noop {} {\bibfield  {journal} {\bibinfo  {journal} {arXiv preprint arXiv:2211.16437}\ } (\bibinfo {year} {2022})}\BibitemShut {NoStop}%
\bibitem [{\citenamefont {Face}\ and\ \citenamefont {Prober}(1987)}]{face1987nucleation}%
  \BibitemOpen
  \bibfield  {author} {\bibinfo {author} {\bibfnamefont {D.~W.}\ \bibnamefont {Face}}\ and\ \bibinfo {author} {\bibfnamefont {D.~E.}\ \bibnamefont {Prober}},\ }\bibfield  {title} {\bibinfo {title} {Nucleation of body-centered-cubic tantalum films with a thin niobium underlayer},\ }\href@noop {} {\bibfield  {journal} {\bibinfo  {journal} {Journal of Vacuum Science \& Technology A: Vacuum, Surfaces, and Films}\ }\textbf {\bibinfo {volume} {5}},\ \bibinfo {pages} {3408} (\bibinfo {year} {1987})}\BibitemShut {NoStop}%
\bibitem [{\citenamefont {Sajovec}\ \emph {et~al.}(1992)\citenamefont {Sajovec}, \citenamefont {Meuffels},\ and\ \citenamefont {Schober}}]{sajovec1992structural}%
  \BibitemOpen
  \bibfield  {author} {\bibinfo {author} {\bibfnamefont {F.}~\bibnamefont {Sajovec}}, \bibinfo {author} {\bibfnamefont {P.~M.}\ \bibnamefont {Meuffels}},\ and\ \bibinfo {author} {\bibfnamefont {T.}~\bibnamefont {Schober}},\ }\bibfield  {title} {\bibinfo {title} {Structural and electrical properties of ion beam sputter deposited tantalum films},\ }\href@noop {} {\bibfield  {journal} {\bibinfo  {journal} {Thin Solid Films}\ }\textbf {\bibinfo {volume} {219}},\ \bibinfo {pages} {206} (\bibinfo {year} {1992})}\BibitemShut {NoStop}%
\bibitem [{\citenamefont {Urade}\ \emph {et~al.}(2024)\citenamefont {Urade}, \citenamefont {Yakushiji}, \citenamefont {Tsujimoto}, \citenamefont {Yamada}, \citenamefont {Makise}, \citenamefont {Mizubayashi},\ and\ \citenamefont {Inomata}}]{urade2024microwave}%
  \BibitemOpen
  \bibfield  {author} {\bibinfo {author} {\bibfnamefont {Y.}~\bibnamefont {Urade}}, \bibinfo {author} {\bibfnamefont {K.}~\bibnamefont {Yakushiji}}, \bibinfo {author} {\bibfnamefont {M.}~\bibnamefont {Tsujimoto}}, \bibinfo {author} {\bibfnamefont {T.}~\bibnamefont {Yamada}}, \bibinfo {author} {\bibfnamefont {K.}~\bibnamefont {Makise}}, \bibinfo {author} {\bibfnamefont {W.}~\bibnamefont {Mizubayashi}},\ and\ \bibinfo {author} {\bibfnamefont {K.}~\bibnamefont {Inomata}},\ }\bibfield  {title} {\bibinfo {title} {Microwave characterization of tantalum superconducting resonators on silicon substrate with niobium buffer layer},\ }\href@noop {} {\bibfield  {journal} {\bibinfo  {journal} {APL Materials}\ }\textbf {\bibinfo {volume} {12}} (\bibinfo {year} {2024})}\BibitemShut {NoStop}%
\bibitem [{\citenamefont {Chen}\ \emph {et~al.}(2001)\citenamefont {Chen}, \citenamefont {Chen}, \citenamefont {Huang},\ and\ \citenamefont {Lee}}]{chen2001growth}%
  \BibitemOpen
  \bibfield  {author} {\bibinfo {author} {\bibfnamefont {G.~S.}\ \bibnamefont {Chen}}, \bibinfo {author} {\bibfnamefont {S.~T.}\ \bibnamefont {Chen}}, \bibinfo {author} {\bibfnamefont {S.~C.}\ \bibnamefont {Huang}},\ and\ \bibinfo {author} {\bibfnamefont {H.~Y.}\ \bibnamefont {Lee}},\ }\bibfield  {title} {\bibinfo {title} {Growth mechanism of sputter deposited ta and ta--n thin films induced by an underlying titanium layer and varying nitrogen flow rates},\ }\href@noop {} {\bibfield  {journal} {\bibinfo  {journal} {Applied surface science}\ }\textbf {\bibinfo {volume} {169}},\ \bibinfo {pages} {353} (\bibinfo {year} {2001})}\BibitemShut {NoStop}%
\bibitem [{\citenamefont {Zhang}\ \emph {et~al.}(2003)\citenamefont {Zhang}, \citenamefont {Huai}, \citenamefont {Chen},\ and\ \citenamefont {Zhang}}]{zhang2003formation}%
  \BibitemOpen
  \bibfield  {author} {\bibinfo {author} {\bibfnamefont {J.}~\bibnamefont {Zhang}}, \bibinfo {author} {\bibfnamefont {Y.}~\bibnamefont {Huai}}, \bibinfo {author} {\bibfnamefont {L.}~\bibnamefont {Chen}},\ and\ \bibinfo {author} {\bibfnamefont {J.}~\bibnamefont {Zhang}},\ }\bibfield  {title} {\bibinfo {title} {Formation of low resistivity alpha ta by ion beam sputtering},\ }\href@noop {} {\bibfield  {journal} {\bibinfo  {journal} {Journal of Vacuum Science \& Technology B: Microelectronics and Nanometer Structures Processing, Measurement, and Phenomena}\ }\textbf {\bibinfo {volume} {21}},\ \bibinfo {pages} {237} (\bibinfo {year} {2003})}\BibitemShut {NoStop}%
\bibitem [{\citenamefont {Wu}\ \emph {et~al.}(2023)\citenamefont {Wu}, \citenamefont {Ding}, \citenamefont {Xiong},\ and\ \citenamefont {Feng}}]{wu2023high}%
  \BibitemOpen
  \bibfield  {author} {\bibinfo {author} {\bibfnamefont {Y.}~\bibnamefont {Wu}}, \bibinfo {author} {\bibfnamefont {Z.}~\bibnamefont {Ding}}, \bibinfo {author} {\bibfnamefont {K.}~\bibnamefont {Xiong}},\ and\ \bibinfo {author} {\bibfnamefont {J.}~\bibnamefont {Feng}},\ }\bibfield  {title} {\bibinfo {title} {High-quality superconducting $\alpha$-ta film sputtered on the heated silicon substrate},\ }\href@noop {} {\bibfield  {journal} {\bibinfo  {journal} {Scientific Reports}\ }\textbf {\bibinfo {volume} {13}},\ \bibinfo {pages} {12810} (\bibinfo {year} {2023})}\BibitemShut {NoStop}%
\bibitem [{\citenamefont {Bernoulli}\ \emph {et~al.}(2013)\citenamefont {Bernoulli}, \citenamefont {M{\"u}ller}, \citenamefont {Schwarzenberger}, \citenamefont {Hauert},\ and\ \citenamefont {Spolenak}}]{bernoulli2013magnetron}%
  \BibitemOpen
  \bibfield  {author} {\bibinfo {author} {\bibfnamefont {D.}~\bibnamefont {Bernoulli}}, \bibinfo {author} {\bibfnamefont {U.}~\bibnamefont {M{\"u}ller}}, \bibinfo {author} {\bibfnamefont {M.}~\bibnamefont {Schwarzenberger}}, \bibinfo {author} {\bibfnamefont {R.}~\bibnamefont {Hauert}},\ and\ \bibinfo {author} {\bibfnamefont {R.}~\bibnamefont {Spolenak}},\ }\bibfield  {title} {\bibinfo {title} {Magnetron sputter deposited tantalum and tantalum nitride thin films: An analysis of phase, hardness and composition},\ }\href@noop {} {\bibfield  {journal} {\bibinfo  {journal} {Thin Solid Films}\ }\textbf {\bibinfo {volume} {548}},\ \bibinfo {pages} {157} (\bibinfo {year} {2013})}\BibitemShut {NoStop}%
\bibitem [{\citenamefont {Gladczuk}\ \emph {et~al.}(2005)\citenamefont {Gladczuk}, \citenamefont {Patel}, \citenamefont {Demaree},\ and\ \citenamefont {Sosnowski}}]{gladczuk2005sputter}%
  \BibitemOpen
  \bibfield  {author} {\bibinfo {author} {\bibfnamefont {L.}~\bibnamefont {Gladczuk}}, \bibinfo {author} {\bibfnamefont {A.}~\bibnamefont {Patel}}, \bibinfo {author} {\bibfnamefont {J.~D.}\ \bibnamefont {Demaree}},\ and\ \bibinfo {author} {\bibfnamefont {M.}~\bibnamefont {Sosnowski}},\ }\bibfield  {title} {\bibinfo {title} {Sputter deposition of bcc tantalum films with tan underlayers for protection of steel},\ }\href@noop {} {\bibfield  {journal} {\bibinfo  {journal} {Thin Solid Films}\ }\textbf {\bibinfo {volume} {476}},\ \bibinfo {pages} {295} (\bibinfo {year} {2005})}\BibitemShut {NoStop}%
\bibitem [{\citenamefont {Wang}\ \emph {et~al.}(2016)\citenamefont {Wang}, \citenamefont {Chen}, \citenamefont {Peng}, \citenamefont {Kuo}, \citenamefont {Yeh}, \citenamefont {Chien},\ and\ \citenamefont {Ying}}]{wang2016influence}%
  \BibitemOpen
  \bibfield  {author} {\bibinfo {author} {\bibfnamefont {W.-L.}\ \bibnamefont {Wang}}, \bibinfo {author} {\bibfnamefont {W.-C.}\ \bibnamefont {Chen}}, \bibinfo {author} {\bibfnamefont {K.-T.}\ \bibnamefont {Peng}}, \bibinfo {author} {\bibfnamefont {H.-C.}\ \bibnamefont {Kuo}}, \bibinfo {author} {\bibfnamefont {M.-H.}\ \bibnamefont {Yeh}}, \bibinfo {author} {\bibfnamefont {H.-J.}\ \bibnamefont {Chien}},\ and\ \bibinfo {author} {\bibfnamefont {T.-H.}\ \bibnamefont {Ying}},\ }\bibfield  {title} {\bibinfo {title} {The influence of amorphous tanx under-layer on the crystal growth of over-deposited ta film},\ }\href@noop {} {\bibfield  {journal} {\bibinfo  {journal} {Thin Solid Films}\ }\textbf {\bibinfo {volume} {603}},\ \bibinfo {pages} {34} (\bibinfo {year} {2016})}\BibitemShut {NoStop}%
\bibitem [{\citenamefont {Tsao}\ \emph {et~al.}(2013)\citenamefont {Tsao}, \citenamefont {Liu}, \citenamefont {Fang},\ and\ \citenamefont {Wang}}]{tsao2013tantalum}%
  \BibitemOpen
  \bibfield  {author} {\bibinfo {author} {\bibfnamefont {J.-C.}\ \bibnamefont {Tsao}}, \bibinfo {author} {\bibfnamefont {C.-P.}\ \bibnamefont {Liu}}, \bibinfo {author} {\bibfnamefont {H.-C.}\ \bibnamefont {Fang}},\ and\ \bibinfo {author} {\bibfnamefont {Y.-L.}\ \bibnamefont {Wang}},\ }\bibfield  {title} {\bibinfo {title} {How tantalum proceeds phase change on tantalum nitride underlayer with sequential ar plasma treatment},\ }\href@noop {} {\bibfield  {journal} {\bibinfo  {journal} {Materials Chemistry and Physics}\ }\textbf {\bibinfo {volume} {137}},\ \bibinfo {pages} {689} (\bibinfo {year} {2013})}\BibitemShut {NoStop}%
\bibitem [{\citenamefont {Alegria}\ \emph {et~al.}(2023)\citenamefont {Alegria}, \citenamefont {Tennant}, \citenamefont {Chaves}, \citenamefont {Lee}, \citenamefont {O'Kelley}, \citenamefont {Rosen},\ and\ \citenamefont {DuBois}}]{alegria2023two}%
  \BibitemOpen
  \bibfield  {author} {\bibinfo {author} {\bibfnamefont {L.~D.}\ \bibnamefont {Alegria}}, \bibinfo {author} {\bibfnamefont {D.~M.}\ \bibnamefont {Tennant}}, \bibinfo {author} {\bibfnamefont {K.~R.}\ \bibnamefont {Chaves}}, \bibinfo {author} {\bibfnamefont {J.~R.~I.}\ \bibnamefont {Lee}}, \bibinfo {author} {\bibfnamefont {S.~R.}\ \bibnamefont {O'Kelley}}, \bibinfo {author} {\bibfnamefont {Y.~J.}\ \bibnamefont {Rosen}},\ and\ \bibinfo {author} {\bibfnamefont {J.~L.}\ \bibnamefont {DuBois}},\ }\bibfield  {title} {\bibinfo {title} {Two-level systems in nucleated and non-nucleated epitaxial alpha-tantalum films},\ }\href@noop {} {\bibfield  {journal} {\bibinfo  {journal} {Applied Physics Letters}\ }\textbf {\bibinfo {volume} {123}} (\bibinfo {year} {2023})}\BibitemShut {NoStop}%
\bibitem [{\citenamefont {M{\"u}ller}\ \emph {et~al.}(2019)\citenamefont {M{\"u}ller}, \citenamefont {Cole},\ and\ \citenamefont {Lisenfeld}}]{muller2019towards}%
  \BibitemOpen
  \bibfield  {author} {\bibinfo {author} {\bibfnamefont {C.}~\bibnamefont {M{\"u}ller}}, \bibinfo {author} {\bibfnamefont {J.~H.}\ \bibnamefont {Cole}},\ and\ \bibinfo {author} {\bibfnamefont {J.}~\bibnamefont {Lisenfeld}},\ }\bibfield  {title} {\bibinfo {title} {Towards understanding two-level-systems in amorphous solids: insights from quantum circuits},\ }\href@noop {} {\bibfield  {journal} {\bibinfo  {journal} {Reports on Progress in Physics}\ }\textbf {\bibinfo {volume} {82}},\ \bibinfo {pages} {124501} (\bibinfo {year} {2019})}\BibitemShut {NoStop}%
\bibitem [{\citenamefont {Kopas}\ \emph {et~al.}(2022)\citenamefont {Kopas}, \citenamefont {Lachman}, \citenamefont {McRae}, \citenamefont {Mohan}, \citenamefont {Mutus}, \citenamefont {Nersisyan},\ and\ \citenamefont {Poudel}}]{kopas2022simple}%
  \BibitemOpen
  \bibfield  {author} {\bibinfo {author} {\bibfnamefont {C.~J.}\ \bibnamefont {Kopas}}, \bibinfo {author} {\bibfnamefont {E.}~\bibnamefont {Lachman}}, \bibinfo {author} {\bibfnamefont {C.~R.~H.}\ \bibnamefont {McRae}}, \bibinfo {author} {\bibfnamefont {Y.}~\bibnamefont {Mohan}}, \bibinfo {author} {\bibfnamefont {J.~Y.}\ \bibnamefont {Mutus}}, \bibinfo {author} {\bibfnamefont {A.}~\bibnamefont {Nersisyan}},\ and\ \bibinfo {author} {\bibfnamefont {A.}~\bibnamefont {Poudel}},\ }\bibfield  {title} {\bibinfo {title} {Simple coplanar waveguide resonator mask targeting metal-substrate interface},\ }\href@noop {} {\bibfield  {journal} {\bibinfo  {journal} {arXiv preprint arXiv:2204.07202}\ } (\bibinfo {year} {2022})}\BibitemShut {NoStop}%
\bibitem [{\citenamefont {Calusine}\ \emph {et~al.}(2018)\citenamefont {Calusine}, \citenamefont {Melville}, \citenamefont {Woods}, \citenamefont {Das}, \citenamefont {Stull}, \citenamefont {Bolkhovsky}, \citenamefont {Braje}, \citenamefont {Hover}, \citenamefont {Kim}, \citenamefont {Miloshi} \emph {et~al.}}]{calusine2018analysis}%
  \BibitemOpen
  \bibfield  {author} {\bibinfo {author} {\bibfnamefont {G.}~\bibnamefont {Calusine}}, \bibinfo {author} {\bibfnamefont {A.}~\bibnamefont {Melville}}, \bibinfo {author} {\bibfnamefont {W.}~\bibnamefont {Woods}}, \bibinfo {author} {\bibfnamefont {R.}~\bibnamefont {Das}}, \bibinfo {author} {\bibfnamefont {C.}~\bibnamefont {Stull}}, \bibinfo {author} {\bibfnamefont {V.}~\bibnamefont {Bolkhovsky}}, \bibinfo {author} {\bibfnamefont {D.}~\bibnamefont {Braje}}, \bibinfo {author} {\bibfnamefont {D.}~\bibnamefont {Hover}}, \bibinfo {author} {\bibfnamefont {D.~K.}\ \bibnamefont {Kim}}, \bibinfo {author} {\bibfnamefont {X.}~\bibnamefont {Miloshi}}, \emph {et~al.},\ }\bibfield  {title} {\bibinfo {title} {Analysis and mitigation of interface losses in trenched superconducting coplanar waveguide resonators},\ }\href@noop {} {\bibfield  {journal} {\bibinfo  {journal} {Applied Physics Letters}\ }\textbf {\bibinfo {volume} {112}} (\bibinfo {year} {2018})}\BibitemShut {NoStop}%
\bibitem [{\citenamefont {McRae}\ \emph {et~al.}(2020)\citenamefont {McRae}, \citenamefont {Wang}, \citenamefont {Gao}, \citenamefont {Vissers}, \citenamefont {Brecht}, \citenamefont {Dunsworth}, \citenamefont {Pappas},\ and\ \citenamefont {Mutus}}]{mcrae2020materials}%
  \BibitemOpen
  \bibfield  {author} {\bibinfo {author} {\bibfnamefont {C.~R.~H.}\ \bibnamefont {McRae}}, \bibinfo {author} {\bibfnamefont {H.}~\bibnamefont {Wang}}, \bibinfo {author} {\bibfnamefont {J.}~\bibnamefont {Gao}}, \bibinfo {author} {\bibfnamefont {M.~R.}\ \bibnamefont {Vissers}}, \bibinfo {author} {\bibfnamefont {T.}~\bibnamefont {Brecht}}, \bibinfo {author} {\bibfnamefont {A.}~\bibnamefont {Dunsworth}}, \bibinfo {author} {\bibfnamefont {D.~P.}\ \bibnamefont {Pappas}},\ and\ \bibinfo {author} {\bibfnamefont {J.}~\bibnamefont {Mutus}},\ }\bibfield  {title} {\bibinfo {title} {Materials loss measurements using superconducting microwave resonators},\ }\href@noop {} {\bibfield  {journal} {\bibinfo  {journal} {Review of Scientific Instruments}\ }\textbf {\bibinfo {volume} {91}} (\bibinfo {year} {2020})}\BibitemShut {NoStop}%
\bibitem [{\citenamefont {Lee}\ \emph {et~al.}(2004)\citenamefont {Lee}, \citenamefont {Doxbeck}, \citenamefont {Mueller}, \citenamefont {Cipollo},\ and\ \citenamefont {Cote}}]{lee2004texture}%
  \BibitemOpen
  \bibfield  {author} {\bibinfo {author} {\bibfnamefont {S.~L.}\ \bibnamefont {Lee}}, \bibinfo {author} {\bibfnamefont {M.}~\bibnamefont {Doxbeck}}, \bibinfo {author} {\bibfnamefont {J.}~\bibnamefont {Mueller}}, \bibinfo {author} {\bibfnamefont {M.}~\bibnamefont {Cipollo}},\ and\ \bibinfo {author} {\bibfnamefont {P.}~\bibnamefont {Cote}},\ }\bibfield  {title} {\bibinfo {title} {Texture, structure and phase transformation in sputter beta tantalum coating},\ }\href@noop {} {\bibfield  {journal} {\bibinfo  {journal} {Surface and Coatings Technology}\ }\textbf {\bibinfo {volume} {177}},\ \bibinfo {pages} {44} (\bibinfo {year} {2004})}\BibitemShut {NoStop}%
\bibitem [{\citenamefont {Catania}\ \emph {et~al.}(1992)\citenamefont {Catania}, \citenamefont {Doyle},\ and\ \citenamefont {Cuomo}}]{catania1992low}%
  \BibitemOpen
  \bibfield  {author} {\bibinfo {author} {\bibfnamefont {P.}~\bibnamefont {Catania}}, \bibinfo {author} {\bibfnamefont {J.~P.}\ \bibnamefont {Doyle}},\ and\ \bibinfo {author} {\bibfnamefont {J.~J.}\ \bibnamefont {Cuomo}},\ }\bibfield  {title} {\bibinfo {title} {Low resistivity body-centered cubic tantalum thin films as diffusion barriers between copper and silicon},\ }\href@noop {} {\bibfield  {journal} {\bibinfo  {journal} {Journal of Vacuum Science \& Technology A: Vacuum, Surfaces, and Films}\ }\textbf {\bibinfo {volume} {10}},\ \bibinfo {pages} {3318} (\bibinfo {year} {1992})}\BibitemShut {NoStop}%
\bibitem [{\citenamefont {Khalil}\ \emph {et~al.}(2012)\citenamefont {Khalil}, \citenamefont {Stoutimore}, \citenamefont {Wellstood},\ and\ \citenamefont {Osborn}}]{khalil2012analysis}%
  \BibitemOpen
  \bibfield  {author} {\bibinfo {author} {\bibfnamefont {M.~S.}\ \bibnamefont {Khalil}}, \bibinfo {author} {\bibfnamefont {M.~J.~A.}\ \bibnamefont {Stoutimore}}, \bibinfo {author} {\bibfnamefont {F.}~\bibnamefont {Wellstood}},\ and\ \bibinfo {author} {\bibfnamefont {K.~D.}\ \bibnamefont {Osborn}},\ }\bibfield  {title} {\bibinfo {title} {An analysis method for asymmetric resonator transmission applied to superconducting devices},\ }\href@noop {} {\bibfield  {journal} {\bibinfo  {journal} {Journal of Applied Physics}\ }\textbf {\bibinfo {volume} {111}} (\bibinfo {year} {2012})}\BibitemShut {NoStop}%
\bibitem [{\citenamefont {Jones}\ \emph {et~al.}(2023)\citenamefont {Jones}, \citenamefont {Materise}, \citenamefont {Leung}, \citenamefont {Weber}, \citenamefont {Isakov}, \citenamefont {Chen}, \citenamefont {Zheng}, \citenamefont {Gyenis}, \citenamefont {Jaeck},\ and\ \citenamefont {McRae}}]{jones2023grain}%
  \BibitemOpen
  \bibfield  {author} {\bibinfo {author} {\bibfnamefont {S.~G.}\ \bibnamefont {Jones}}, \bibinfo {author} {\bibfnamefont {N.}~\bibnamefont {Materise}}, \bibinfo {author} {\bibfnamefont {K.~W.}\ \bibnamefont {Leung}}, \bibinfo {author} {\bibfnamefont {J.~C.}\ \bibnamefont {Weber}}, \bibinfo {author} {\bibfnamefont {B.~D.}\ \bibnamefont {Isakov}}, \bibinfo {author} {\bibfnamefont {X.}~\bibnamefont {Chen}}, \bibinfo {author} {\bibfnamefont {J.}~\bibnamefont {Zheng}}, \bibinfo {author} {\bibfnamefont {A.}~\bibnamefont {Gyenis}}, \bibinfo {author} {\bibfnamefont {B.}~\bibnamefont {Jaeck}},\ and\ \bibinfo {author} {\bibfnamefont {C.~R.~H.}\ \bibnamefont {McRae}},\ }\bibfield  {title} {\bibinfo {title} {Grain size in low loss superconducting ta thin films on c axis sapphire},\ }\href@noop {} {\bibfield  {journal} {\bibinfo  {journal} {Journal of Applied Physics}\ }\textbf {\bibinfo {volume} {134}} (\bibinfo {year} {2023})}\BibitemShut {NoStop}%
\bibitem [{\citenamefont {Lutchyn}\ \emph {et~al.}(2018)\citenamefont {Lutchyn}, \citenamefont {Bakkers}, \citenamefont {Kouwenhoven}, \citenamefont {Krogstrup}, \citenamefont {Marcus},\ and\ \citenamefont {Oreg}}]{lutchyn2018majorana}%
  \BibitemOpen
  \bibfield  {author} {\bibinfo {author} {\bibfnamefont {R.~M.}\ \bibnamefont {Lutchyn}}, \bibinfo {author} {\bibfnamefont {E.~P. A.~M.}\ \bibnamefont {Bakkers}}, \bibinfo {author} {\bibfnamefont {L.~P.}\ \bibnamefont {Kouwenhoven}}, \bibinfo {author} {\bibfnamefont {P.}~\bibnamefont {Krogstrup}}, \bibinfo {author} {\bibfnamefont {C.~M.}\ \bibnamefont {Marcus}},\ and\ \bibinfo {author} {\bibfnamefont {Y.}~\bibnamefont {Oreg}},\ }\bibfield  {title} {\bibinfo {title} {Majorana zero modes in superconductor--semiconductor heterostructures},\ }\href@noop {} {\bibfield  {journal} {\bibinfo  {journal} {Nature Reviews Materials}\ }\textbf {\bibinfo {volume} {3}},\ \bibinfo {pages} {52} (\bibinfo {year} {2018})}\BibitemShut {NoStop}%
\end{thebibliography}%

\end{document}